\documentclass[10pt]{article}

\usepackage{authblk}

\usepackage{graphicx}
\usepackage{amsmath,amsthm,bm,mathrsfs,subfloat}
\usepackage{appendix}
\usepackage[normalem]{ulem}
\usepackage{comment}
\usepackage{cite}
\usepackage{xcolor,paralist,hyperref,titlesec,fancyhdr,etoolbox}
\usepackage{geometry}
\title{Elastic jump propagation across a blood vessel junction}
\author[1]{Tamsin A.~Spelman}
\author[2]{Ifeanyi S.~Onah}
\author[2]{David MacTaggart}
\author[2]{Peter S.~Stewart}
\affil[1]{Sainsbury Laboratory, University of Cambridge, 47 Bateman Street, Cambridge, CB2 1LR, UK}
\affil[2]{School of Mathematics and Statistics, University of Glasgow, University Place, Glasgow, G12 8SQ, UK}

\begin{document}
\maketitle

\begin{abstract} 
{The theory of small-amplitude waves propagating across a blood vessel junction has been well established with linear analysis. In this study we consider the propagation of large-amplitude, nonlinear waves (\emph{i.e.} shocks and rarefactions) through a junction from a parent vessel into two (identical) daughter vessels using a combination of three approaches: numerical computations using a Godunov method with patching across the junction, analysis of a nonlinear Riemann problem in the neighbourhood of the junction and an analytical theory which extends the linear analysis to the following order in amplitude. A unified picture emerges: an abrupt (prescribed) increase in pressure at the inlet to the parent vessel generates a propagating shock wave along the parent vessel which interacts with the junction. For modest driving, this shock wave divides into propagating shock waves along the two daughter vessels and reflects a rarefaction wave back towards the inlet. However, for larger driving the reflected rarefaction wave becomes transcritical, generating an additional shock wave. Just beyond criticality this new shock wave has zero speed, pinned to the junction, but for further increases in driving this additional shock divides into two new propagating shock waves in the daughter vessels. }
\end{abstract}

\section{Introduction}

The human circulatory system is formed of a bifurcating network of arteries and veins which conduct blood to and from the heart. Each heartbeat drives a rapid pulse wave through the system, which is easily detected in the larger arteries as a non-invasive clinical measure of cardiovascular health. Over the years there have been a wide variety of theoretical models to characterise pulse propagation though the cardiovascular system. Early models considered the propagation of small-amplitude pressure waves through a bifurcating network of elastic-walled tubes conveying fluid (see summaries in the monographs of Lighthill \cite{Lighthill_Book} and Pedley \cite{Pedleybook}). The inviscid flow in each constituent vessel is described through a characteristic impedance and the division of mass across each junction is characterised by a reflection coefficient; this simple description facilitates coupling into small networks \cite{taylor1966input}. Such approaches have since been extended to larger networks to mimic the human cardiovascular system with physiological parameter values (\emph{e.g.} \cite{avolio1980multi,alastruey2009analysing}). This linearised analysis is typically sufficient to capture the dynamics because the flow remains strongly subcritical, where the maximal Mach number (the ratio of the maximal flow speed to the wavespeed) has been estimated as $0.2-0.25$ in dogs, with humans believed to be similar \cite{Pedleybook}.

However, abnormal physiological events can also trigger pressure wave propagation through the cardiovascular system. For example, incompetent aortic values can increase the volume of blood ejected by the heart, significantly enlarging the amplitude of the pressure pulse (sometimes known as a `pistol shot' pulse) \cite{beck1935aortic,lange1956observations}. Theoretical modelling of the propagation of these large-amplitude pressure waves is significantly more involved than the linearised analysis discussed above, where the models must now account for wave steepening and shock (or elastic jump) formation \cite{anliker1971nonlinear}. The theory of shock waves propagating along a single elastic-walled blood vessel is similar to other classical hyperbolic systems, such as the dam break problem in shallow water theory \cite{whitham,ockendon2004waves}. In particular, classical analysis using the tools of nonlinear waves \cite{whitham,lighthill} can predict the distance along the aorta from the heart where a shock will first form \cite{Pedleybook}. 

Large-amplitude pressure waves can also propagate as a result of external events. The present study is motivated by the aim to quantify the onset of retinal haemorrhage (\emph{i.e.} bleeding of the retinal blood vessels at the back of the eye) following a traumatic brain injury, particularly in infants \cite{levin2009retinal}. One hypothesised mechanism is that retinal haemorrhage arises as a result of an abrupt increase in intracranial pressure (the so-called `pressure-rise hypothesis') \cite{shiau2012retinal} which is then transmitted to the cerebrospinal fluid within the meningeal sheath of the optic nerve (the large cable of dendritic fibres connecting the eye to the brain) \cite{medele1998terson,ko2010ocular}. Behind the eye, modelling predicts that such a pressure increase can drive a propagating shock wave along the nerve sheath, which is then significantly amplified by reflection at the blind end at the sclera and which can rupture bridging blood vessels and lead to optic nerve haemorrhage \cite{bonshek2014study,stewart2023model}. However, the central retinal artery and vein traverse the optic nerve sheath as they enter the eye, and so are exposed to this acute increase in external pressure, which they then transmit to the retinal circulation. Modelling predicts that the amplitude and shape of the pressure wave transmitted into the retina is linked to the length of vessel confined within the optic nerve just before it enters the eye; for large enough perturbations this pressure wave can steepen to form a shock in the retinal circulation, where the tail of this incoming pressure wave is distributed over such long lengthscales that it can be approximated as spatially uniform \cite{Spelman2020}. The clinical hypothesis attributes retinal haemorrhage to the propagation of this pressure wave, which will spread through the retinal circulation and drive abrupt expansion, and possible rupture, of more fragile vessels in distal parts of the network. Hence, an understanding of how large-amplitude waves propagate through blood vessel junctions is an essential component of any predictive model for retinal haemorrhage. As a precursor to developing large-scale computational models for the retinal circulation in response to a large-amplitude pressure perturbation, this paper focuses specifically on the propagation of a large-amplitude disturbance across a single blood vessel junction.

Within each individual blood vessel, the governing equations for incompressible blood flow are typically hyperbolic. Three-dimensional fluid-structure interaction models require careful numerical treatment to resolve the wave propagation (\emph{e.g.} \cite{chabannes2013high,kafi2016numerical,jin2021arterial}). Such fluid-structure interaction approaches can be extended to consider three-dimensional flows through single blood vessel junctions (\emph{e.g.} \cite{dewilde2016assessment}) but extensions to model large portions of the cardiovascular system in this way are computationally demanding, particularly for networks composed of multiple generations (\emph{e.g.} \cite{moerman2022development}). A more computationally tractable approach is to cross-sectionally average the three-dimensional governing equations and use a closure condition to derive a spatially one-dimensional representation of the flow. Even these reduced equations can present a significant numerical challenge to resolve the wave propagation, and are typically solved numerically using bespoke finite volume schemes (\emph{e.g.} \cite{olufsen2000numerical,2003_Sherwin}). Such models have been used to investigate additional effects such as vessel tapering (\emph{e.g.} \cite{abdullateef2021impact}), wall viscoelasticity (\emph{e.g.} \cite{muller2016consistent,piccioli2022modeling}) and blood rheology (\emph{e.g.} \cite{ghigo2018time}).

These lower-order models for individual vessels can be extended to networks by coupling spatially one-dimensional flow within each vessels through boundary conditions imposed at single points which mimic the junction. These models impose conservation of (total or static) pressure and flow, either by directly imposing continuity across the junction using extrapolation along characteristics to inform the matching (\emph{e.g.} \cite{olufsen2000numerical,2003_Sherwin,reymond2009validation,muller2014global}) or by introducing a separate junction compartment (\emph{e.g.} \cite{Zhang2017,2009_Fullana, bellamoli2018numerical}). 

When the Reynolds number is large enough, these hyperbolic systems can exhibit wave steepening and thus the numerical schemes must be specifically developed to allow the propagation of shock waves. The seminal work on shock capture in blood vessels was conducted by Brook \emph{et al.} \cite{Brook1999}, who developed a numerical scheme which exploits a local Riemann problem in each finite volume; this Riemann problem has subsequently been studied in detail using sophisticated constitutive laws for the vessel wall mechanics for both arteries \cite{toro2012simplified} and veins \cite{spiller2017exact}. To extend this approach to circulatory networks the Riemann problem first needs to be adapted for use across a junction, known as the junction Riemann problem \cite{contarino2016junction,Murillo2021}. In this case the junction boundary conditions are embedded at the point of discontinuity in initial conditions of the Riemann problem, similar to an abrupt discontinuity in vessel stiffness \cite{toro2012simplified,han2015subcritical}. The analysis is analogous to these classical Riemann problems while the flow remains subcritical, but extension to transcritical and supercritical flows is not straightforward \cite{contarino2016junction}. Subsequent analysis has posed iterative methods to deal with the supercritical regime \cite{Murillo2021}, but what is currently missing is an understanding of how the flow can become transcritical, as may occur in physiological situations such as when a vein becomes highly collapsed \cite{contarino2016junction}. In the analysis below we extend the junction Riemann problem to include transcritical flow, which we show spans a wide range of the parameter space.

In this paper, we explore highly nonlinear wave propagation across a single blood vessel junction composed of a parent vessel bifurcating into two identical daughter vessels. For simplicity, the imposed pressure forcing at the inlet of the parent vessel mimics our prediction of the pressure wave transmitted into the retinal circulation following an abrupt rise in intracranial pressure \cite{Spelman2020}. Within this setup, we revisit the junction Riemann problem and extend the existing analysis to include transcritical flow, demonstrating how resonance can produce additional shock waves in the system. These are similar in spirit to the resonances observed by Han \emph{et al.} who considered the Riemann problem in a single blood vessel with an abrupt discontinuity in elastic stiffness \cite{han2015subcritical,han2015supercritical,onah2023thesis} and analogous resonances found in shallow water theory \cite{lefloch2007riemann,lefloch2011godunov}.

The paper is arranged as follows. In Sec.~\ref{sec:model} we summarise our numerical method for modelling the junction as well as a separate second-order analytical approximation, a continuation of the classical small-amplitude approach to the following order in perturbation amplitude. Predictions of this model are presented in Sec.~\ref{sec:results}, along with evidence of a limit point where the behaviour of the system changes fundamentally. In Sec.~\ref{sec:riemann} we present our version of the junction Riemann problem, which we compare against our fully numerical model. Finally, in Sec.~\ref{sec:stiffness} we consider a vessel bifurcation with a jump in elastic stiffness, and in Sec.~\ref{sec:baseflow} we consider the effect of underlying flow through the junction. We conclude with a discussion and an outline of further work in Sec.~\ref{sec:discussion}.

\section{Model}
\label{sec:model}

\begin{figure}
\centering
\includegraphics[width=6in]{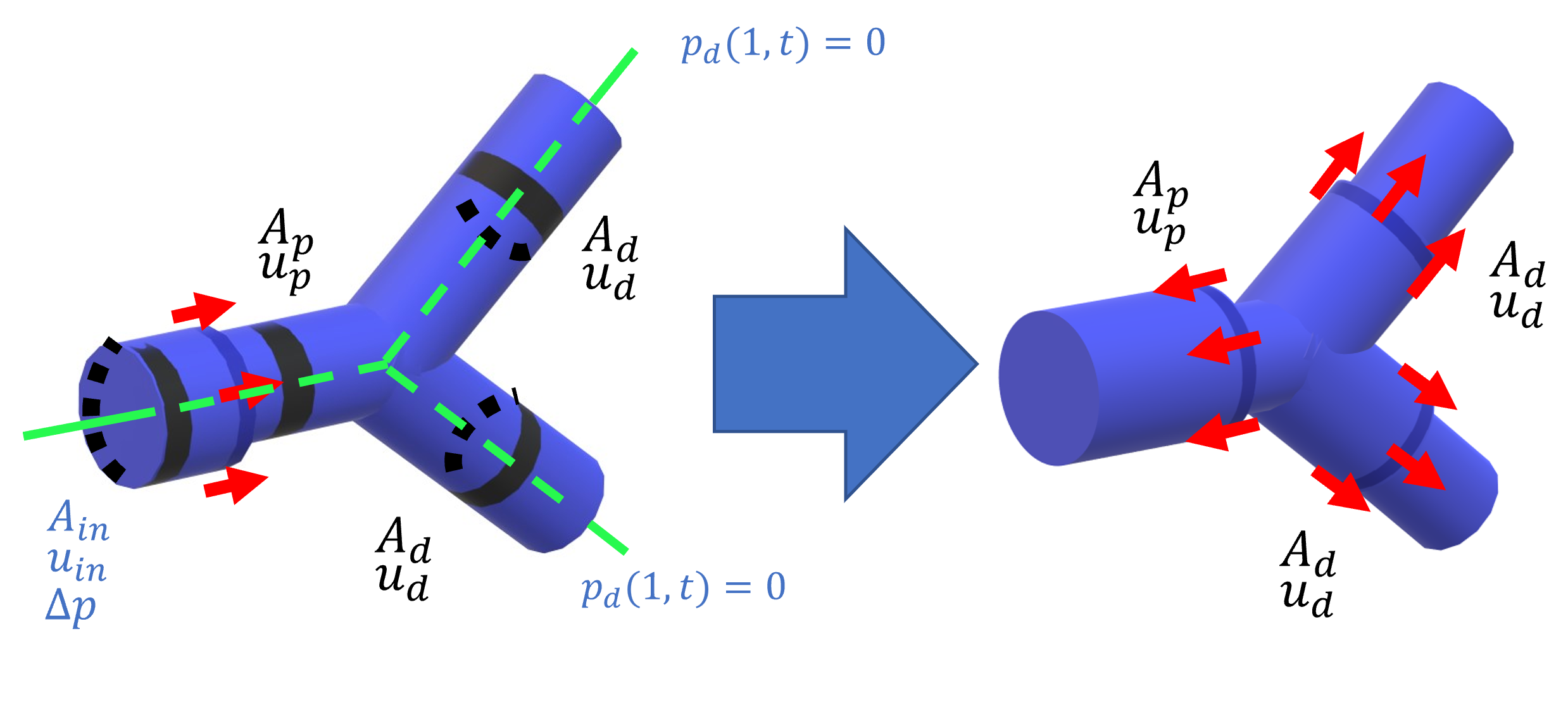}
\caption{Setup of the mathematical model with a parent blood vessel dividing into two identical daughter vessels: (left panel) a prescribed driving pressure applied at the proximal end of the parent vessel initiates wave propagation; (right panel) subsequent transmission of the pressure wave into the daughter vessels accompanied by partial reflection back along the parent vessel.}
\label{fig_Setup}
\end{figure}

Our model focuses on the propagation of a large-amplitude pressure wave through a single blood vessel junction which bifurcates symmetrically into two identical daughter vessels (see Fig.~\ref{fig_Setup}). In this study we ignore the complicated three-dimensional flows local to the junction induced by the geometry of the bifurcation, but instead focus on the global properties of the system induced by splitting the flow leaving the parent vessel and moving into the daughter vessels, as a pre-cursor to studying flow in large-scale cardiovascular networks.

Each constituent vessel is modelled as a long elastic-walled tube of length $L^\ast_i$ ($i=p,d$), where subscripts $i={p,d}$ refer to quantities in the parent or daughter vessels, respectively.  Note that the two daughter vessels are identical, so the single subscript $d$ describes both simultaneously. Each vessel is parameterised by the spatial coordinate $x^*$ along the tube centre line. The junction is denoted as $x^\ast=0$, and so the parent vessel occupies $-L_p^\ast \le x^\ast \le 0$ while the (identical) daughter vessels occupy $0 \le x^\ast \le L_d^\ast$. We systematically reduce the complexity of the three-dimensional flow along the constituent vessels, resulting in a model that has been used in many previous studies of blood flow in flexible-walled vessels \emph{e.g.} \cite{Pedleybook,Brook1999,olufsen2000numerical,muller2013well}, with a full derivation given in our previous study \cite{Spelman2020}. The reduced flow equations are then spatially one-dimensional, where at time $t$ the blood flow in each constituent vessel can be described by three quantities: fluid velocity along the axis of the vessel, $u_{i}^*(x^*,t^*)$; the tube cross-sectional area $A_{i}^*(x^*,t^*)$ and the local blood pressure $p_i^\ast(x^\ast,t^\ast)$, where $i={p,d}$. Furthermore, each vessel has a (uniform) baseline cross-sectional area denoted $\overline{A}^*_i$ ($i=p,d$) and all vessels are subject to a uniform external pressure $p^\ast_e$. We characterise the elasticity of each vessel wall through an elastic stiffness parameter $k^*_i$ ($i=p,d$), which can be estimated from the Young's modulus of the tissue and the wall thickness (see details in our previous study \cite{Spelman2020}). 

The blood within each vessel is modelled as a homogeneous fluid of constant density $\rho$. Blood flow in the retinal circulation is viscous, but here the viscosity of the blood is neglected for two reasons. Firstly, this is the simplest version of the model which admits the formation of large-amplitude nonlinear waves (shocks and rarefactions). Secondly, the timescales of traumatic brain injury are typically so fast that viscous effects do not make a significant contribution (\emph{e.g.} impact timescales for falls from altitude are on the order of milliseconds \cite{1963_Snyder}). It would be straightforward to incorporate the effects of viscosity in the governing equations using an empirical term \cite{cancelli1985separated,Brook1999,Spelman2020}.

In this paper, the system is forced by a prescribed pressure profile at the inlet to the parent vessel with amplitude $\Delta p^\ast$, to mimic pressure wave transmission from the proximal end of the network (\emph{e.g.} as a result of the transmission from acute fluid pressure increases in the optic nerve sheath \cite{Spelman2020}). For simplicity, at the outlet to the two daughter vessels a uniform fluid pressure is imposed by setting $p^\ast_d(x^\ast=L^\ast,t^\ast)=p_0^\ast$. 

The governing equations are non-dimensionalised by scaling relative to the properties of the parent vessel: we scale all lengths on the square root of the baseline cross-sectional area in the parent $(\overline{A}^*_p)^{1/2}$, velocities on $u^\ast_0 = (k^*_{p}/\rho)^{1/2}$, time on $t^\ast_0=L^*_p/u^*_0$ and pressures according to $p^* = k^*_p {p}+p^*_0$. Note that dimensionless variables take the same symbol as their dimensional counterpart without the superscript $^\ast$. This results in the following dimensionless groups for $i=p,d$
\begin{equation}
\beta = \frac{{(\overline{A}^*_p)}^{1/2}}{L_p^\ast}, \,\,\,   \Delta p= \frac{\Delta p^*}{k^*_p}, \,\,\, p_{e} = \frac{p_e^\ast-p_0^\ast}{k^\ast}\,\,\,  k_i=\frac{k^*_i}{k^*_p}, \,\,\,  L_i=\frac{L^*_i}{L^*_p}, \,\,\, \overline{A}_i=\frac{\overline{A}^*_i}{\overline{A}^*_p},
\end{equation}
denoting the aspect ratio of the parent tube, the dimensionless driving pressure, the dimensionless external pressure on the system, the dimensionless stiffness of each constituent vessel, the dimensionless length of each constituent vessel and the dimensionless baseline cross-sectional area of each vessel. Note that this choice of non-dimensionalisation gives $k_p=1$ and $\overline{A}_p=1$, and in this study we restrict attention to the case where the parent and daughter vessels have equal base cross-sectional area (\emph{i.e.} $\overline{A}_d=1$).

The governing equations of mass and momentum for each vessel, in the ideal limit as the Reynolds number approaches infinity, are given by 
\begin{align}
 \frac{\partial A_i}{\partial t}+\frac{\partial q_i}{\partial x} &=0,\label{eq:GovEqu_Mass} \\ 
 \frac{\partial q_i}{\partial t}+\frac{\partial}{\partial x}\left(\frac{q_i^2}{A_i} \right)&= A_i\frac{\partial p_i}{\partial x},
\label{eq:GovEqu_Momentum}
\end{align}
where $q_i = u_i A_i$ is the fluid flux through each cross-section of the vessel ($i=p,d$). Note that the governing equations in the parent vessel hold on the domain $-L_p \le x \le 0$, while in the daughter vessel they hold on $0 \le x \le L_d$. The system is closed using a constitutive law (or `tube law') to mimic the wall elasticity in the form
\begin{align}
p_i - p_e = & k_i \Gamma(A_i), \qquad (i=p,d) \label{eq:GovEqu_Pressure}
\end{align}
where $\Gamma$ is a dimensionless function linking the cross-sectional area of each vessel to the local blood pressure \cite{Pedleybook,Brook1999}. Specifically, we use a nonlinear tube law for the wall mechanics in the form
\begin{equation}
\label{eq:tubelaw}
\Gamma(\alpha)= \alpha^{m}-\alpha^{-n},
\end{equation}
where $m,n \ge 0$ are parameters. This tube law is an algebraic function which resists both vessel expansion (for $m>0$) and collapse (for $n>0$). The parameters $m$ and $n$ can be chosen based on experimental measurements, although most data is obtained \emph{ex vivo}. Here, two particular choices are used to illustrate our results: a nonlinear tube law where $m=10$ and $n=-3/2$ (values proposed for the giraffe jugular vein \cite{Brook1999} and often used for characterising human veins \cite{muller2014global,Spelman2020}) and a simple linear tube law where $m=1$ and $n=0$. 

This choice of tube law sets the nonlinear wave speed $c_i$ along each constituent vessel as 
\begin{equation}
c_i^2(x,t) = {A_i \frac{\partial \Gamma(A_i)}{\partial A_i}}= mA_i^{m}+nA_i^{-n}. \qquad \qquad (i=p,d). \label{Elastic_Wavespeed}
\end{equation}
This, in turn, is used to calculate the local Mach number, $M_i(x,t)$, at every spatial location along the vessel, a classical metric for describing transcritical flow in the form
\begin{equation}
M_i(x,t)=\frac{u_i(x,t)}{c_i(x,t)}, \qquad (i=p,d).
\end{equation}

Our approach uses a one-dimensional model to describe the behaviour within each constituent vessel (Eqs.~(\ref{eq:GovEqu_Mass}-\ref{eq:GovEqu_Pressure})), similar to many previous studies of cardiovascular flow. However, boundary conditions across the junction are required to couple these vessels together. In this case we impose conservation of mass and conservation of total pressure (\emph{e.g.} \cite{2003_Sherwin}). Since the daughter vessels are identical it is assumed that mass of fluid leaving the parent is divided equally between them, which then gives two constraints
\begin{align}
r_p A_p(0,t) u_p(0,t)&= r_d A_d(0,t) u_d(0,t), \label{Eq_Junc_Mass}\\
p_p(A_p(0,t)+\tfrac{1}{2}u^2_p(0,t)&=p_d(A_d(0,t))+\tfrac{1}{2}u^2_d(0,t), \label{Eq_Junc_TotPres} 
\end{align}
where the parameter $r_i$ ($i=p,d$) represents the partition of mass across the junction so that $r_p=1$ and $r_d=2$; the choice of these constants can easily be generalised to capture junctions involving more vessels.

We assume that the vessel cross-sectional area, pressure and flow velocity are spatially uniform in each constituent vessel before the pressure perturbation is applied at the upstream end. In most examples considered in this paper, in the initial configuration the vessels are all assumed to take their baseline cross-sectional area $A_i(x,0)=\bar A_{i}=1$ with no flow $u_i(x,0)=0$ ($i=p,d$), so that $p_i(x,0)=0$. These initial conditions automatically satisfy the junction boundary conditions (\ref{Eq_Junc_Mass}-\ref{Eq_Junc_TotPres}).

However, since arteries and veins have baseline flows in opposite directions, in Sec.~\ref{sec:baseflow} below we consider initial conditions with underlying flow. To ensure this initial configuration is an equilibrium state it must obey the boundary conditions at the junction (\ref{Eq_Junc_Mass}-\ref{Eq_Junc_TotPres}). We prescribe the (spatially uniform) flow velocity in the daughter vessels as $\overline{u}_d(x,0)$, which implies that $\overline{u}_p(x,0)= (r_d/r_p) \overline{u}_d(x,0)$ (given that in this study the baseline cross-sectional area of all constituent vessels are assumed equal). Note that this baseline flow must be driven by a prescribed upstream flux as it carries no pressure gradient in this inviscid model.

For simplicity, in this study we fix the domain lengths as $L_p=L_d=1$ and set the external pressure as $p_e=0$.

In summary, this model is a simplified representation of blood flow across a single dichotomous junction, designed to capture the main features associated with large-amplitude wave propagation, and has been kept deliberately simple so we can focus on the dominant physical mechanisms and compare to the classical work involving small amplitude perturbations \cite{Lighthill_Book,Pedleybook}.

In what follows we characterise the system using a variety of measures, but focus particularly on the response of the system adjacent to the junction. We use a superscript $^{(j)}$ to denote a quantity evaluated at the junction when approaching from a given direction. For example, we denote $A_p^{(j)}(t)=A_p(x\rightarrow 0^-,t)$ and $A_d^{(j)}(t)=A_d(x\rightarrow 0^+,t)$.

We solve this theoretical model using three complementary approaches. Our primary approach is to use a numerical method based on the approach of Brook \emph{et al.} \cite{Brook1999}, described in Sec.~\ref{sec:model}\ref{subsec:numerical}. We further use an analytical approach which generalises the small-amplitude (linear) models described above, which is summarised in Sec.~\ref{sec:lin} and Appendix \ref{App_PertTheory}. Finally we consider our version of the junction Riemann problem in Sec.~\ref{sec:riemann}.

\subsection{Finite-volume numerical method}
\label{subsec:numerical}

Our primary method for solving this system involves a finite volume numerical method. For numerical convenience in these simulations the inlet pressure is abruptly ramped over a (short) timescale $t_{ap}$ (non-dimensionalised on $t_0^\ast$) from the baseline value ($p_p=0$) to the driving pressure $p_p=\Delta p$ after which it remains constant, so that
\begin{equation}
\label{eq:inlet}
p(-L_p,t)=\begin{cases}
\Delta p\sin^{2}\left(\frac{\pi t}{2 t_{ap}}\right) & t<t_{ap},\\
\Delta p & t\ge t_{ap}.
\end{cases}
\end{equation}
The system of mass and momentum equations (\ref{eq:GovEqu_Mass}-\ref{eq:GovEqu_Momentum}) is solved numerically using a modified nonlinear Godunov method, where the spatial domains in the parent vessel ($-L_p \le x \le 0$) and the daughter vessels ($0\le x \le L_d$) are discretised onto a uniform mesh forming a line of finite volumes arranged in series. This method was first developed to describe unsteady flow in the giraffe jugular vein \cite{Brook1999} and has subsequently been modified to analyse a line of vessels joined in series \cite{Spelman2020}. At each time step, we calculate the cross-sectional area at the inlet of the parent vessel using the prescribed inlet pressure (\ref{eq:inlet}) via the tube law (\ref{eq:GovEqu_Pressure}) and then compute the corresponding inlet flow velocity $u_p(x=-L_p,t)$ by extrapolating the backward characteristics in the parent vessel assuming the flow is undisturbed far ahead. We proceed by solving a nonlinear Riemann problem for the cross-sectional area and flow velocity in each finite volume. Note that the final mesh point in the parent vessel and the first mesh point in the daughter vessels are coincident at the junction $(x=0)$, though the corresponding flow variables can be different. We couple these variables by extrapolating the characteristics leaving each vessel and applying the junction boundary conditions (\ref{Eq_Junc_Mass}-\ref{Eq_Junc_TotPres}). For simplicity, zero pressure boundary conditions are imposed at the downstream end of the daughter vessels, but these have negligible effect on the predictions since perturbations do not reach the outlet within the simulation time.

We show below that this characteristic extrapolation technique is less robust at larger driving pressures where the rarefaction flow in the parent vessel becomes transcritical. In Sec.~\ref{sec:riemann}, we construct a nonlinear Riemann solver which explicitly accounts for the division of mass and any jump in material properties across the junction, showing that these predictions exhibit strong agreement with our numerical method. We will also compare our results at lower amplitudes to a modified linear theory described in the following section. 

\subsection{Modified linear perturbation theory}
\label{sec:lin}

Our second approach to study this system is a modification of classical linear theory, which has been shown to approximate small amplitude waves propagating across a blood vessel junction \cite{Pedleybook}. To improve this approximation for waves of slightly larger amplitude (the focus of this paper), in Appendix \ref{App_PertTheory} this linear theory (for the linear tube law with $m=0$ and $n=1$) is extended to the following order in perturbation amplitude, calculating the (steady) amplitudes of the constituent waves. In response to an inlet perturbation to the cross-sectional area of amplitude $\Delta A$ (where $\Delta A \ll 1$) so that the cross-sectional area at the inlet of the parent vessel takes the form $A_{in}=A_p(-L_1,t) \approx 1 + \Delta A$, this second-order theory predicts the amplitude of the reflected waves in the parent vessel adjacent to the junction as
\begin{subequations}
\label{MLP}
\begin{align}
A^{(j)}_p &= 1 +  \frac{2\sqrt{k_d}}{2+\sqrt{k_d}}\Delta A  - \frac{12}{(2+\sqrt{k_d})^3}\Delta A^2 + O(\Delta A^3), \label{MLP1}  \\ 
q^{(j)}_p &= 0 + \frac{4}{2+\sqrt{k_d}} \Delta A  + \frac{12}{(2+\sqrt{k_d})^3}\Delta A^2 + O(\Delta A^3), \label{MLP2} 
\end{align}
and the amplitude of the transmitted waves in the daughter vessels adjacent to the junction as
\begin{align}
A^{(j)}_d &=1 +  \frac{2}{\sqrt{k_d}(2+\sqrt{k_d})}\Delta A  + \frac{6}{\sqrt{k_d}(2+\sqrt{k_d})^3}\Delta A^2 + O(\Delta A^3), \label{MLP3} \\ 
q^{(j)}_d &=0 +  \frac{2}{2+\sqrt{k_d}}\Delta A  + \frac{6}{(2+\sqrt{k_d})^3}\Delta A^2 + O(\Delta A^3).   \label{MLP4}
\end{align}
\end{subequations}

\section{Baseline results}
\label{sec:results}

\begin{figure}
\centering
\includegraphics[width=6.5in]{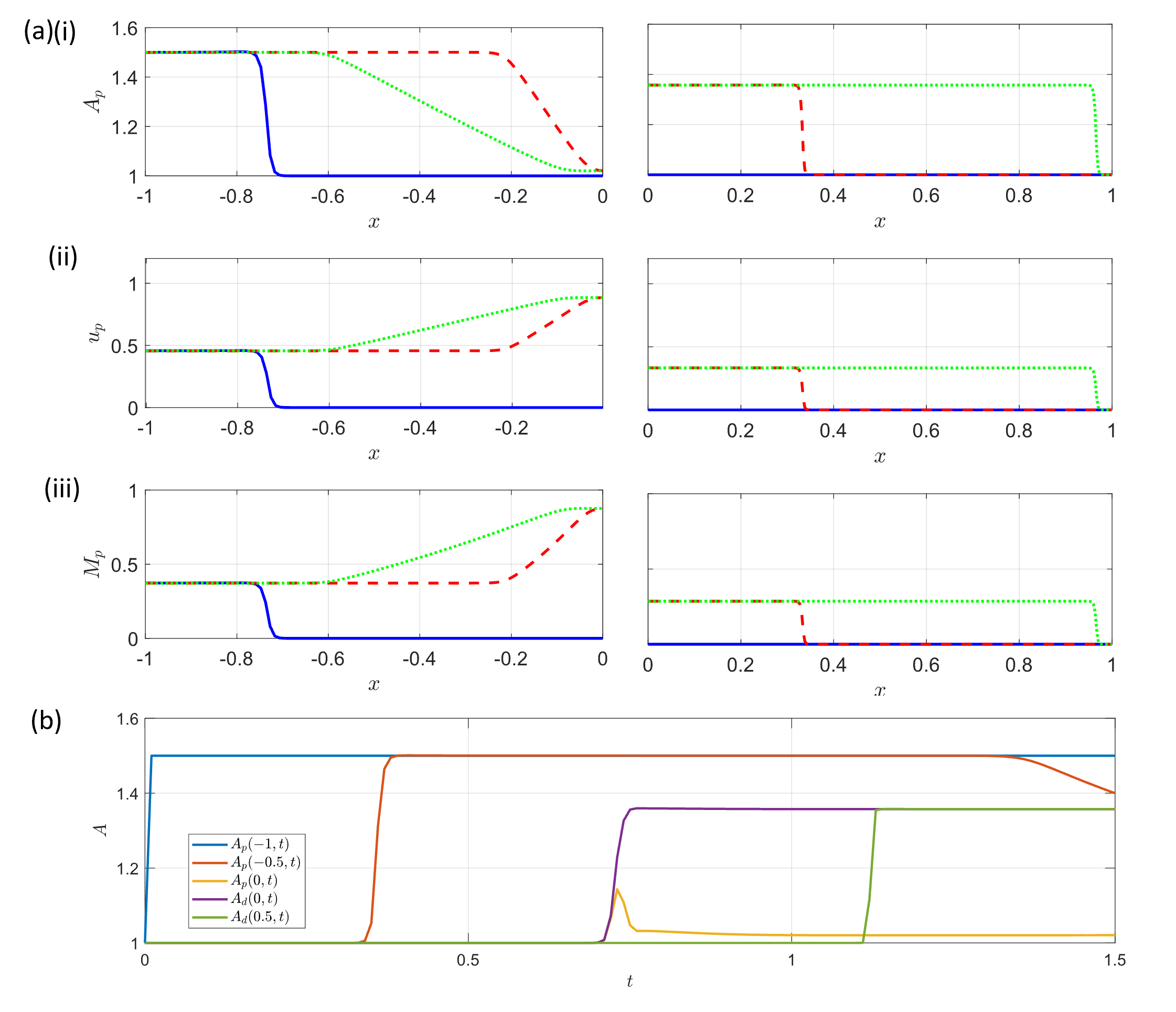}
\caption{Spatial and temporal traces of the junction vessels using the full nonlinear model with a linear tube law $m=1$, $n=0$ in response to a pressure perturbation of amplitude $\Delta p=0.5$. (a) spatial profiles in the parent vessel (left) and daughter vessels (right) at $t=0.19$ (blue solid line), $t=0.99$ (red dotted line) and $t=1.49$ (green double dotted line), for (i) vessel cross-sectional area, (ii) flow velocity and (iii) Mach number; (b) the corresponding time-traces of the vessel cross-sectional area at three spatial locations along the parent vessel (the inlet, $x=-L=-1$, shown as a blue line, the midpoint, $x=-0.5$, shown as a red line, and adjacent to the junction, $x=0^-$, shown as purple line) and two spatial locations in the daughter vessels (adjacent to the junction, $x=0^+$, shown as orange line, and at the midpoint, $x=0.5$, shown as green line). The spatial and temporal dynamics for this case are also shown in the accompanying video.}
\label{fig:Fig2_LinearProfile}
\end{figure}

\begin{figure}
\centering
\includegraphics[width=6.5in]{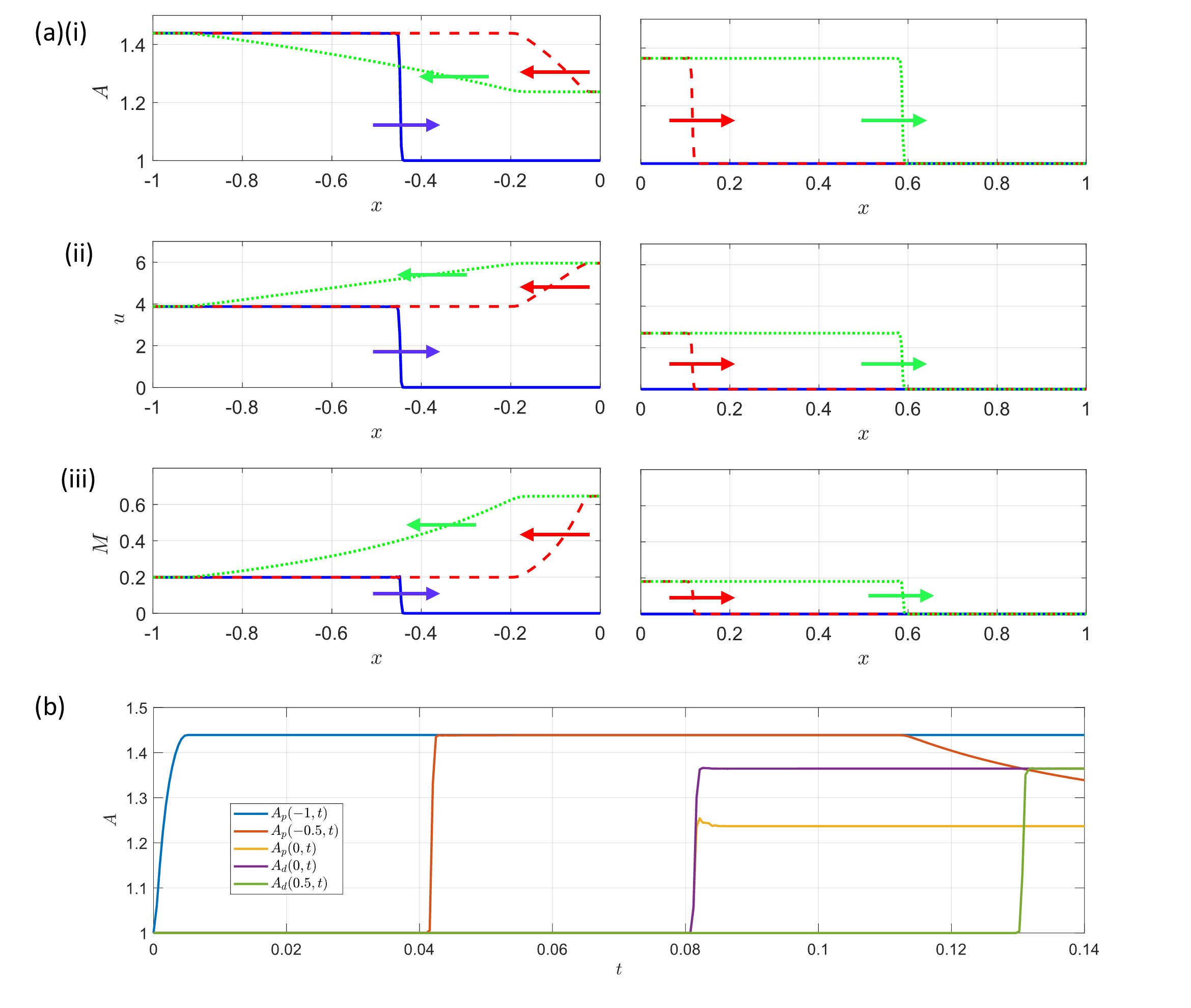}
\caption{Spatial and temporal traces of the junction vessels using the full nonlinear model with a nonlinear tube law $m=10$, $n=3/2$ in response to a pressure perturbation of amplitude $\Delta p=37.5$. (a) spatial profiles in the parent vessel (left) and daughter vessels (right) at $t=0.046$ (blue solid line), $t=0.093$ (red dotted line) and $t=0.14$ (green double dotted line), for (i) vessel cross-sectional area, (ii) flow velocity and (iii) Mach number; (b) the corresponding time-traces of the vessel cross-sectional area at three spatial locations along the parent vessel (the inlet, $x=-L=-1$, shown as a blue line, the midpoint, $x=-0.5$, shown as a red line, and adjacent to the junction, $x=0^-$, shown as purple line) and two spatial locations in the daughter vessels (adjacent to the junction, $x=0^+$, shown as orange line, and at the midpoint, $x=0.5$, shown as green line). The spatial and temporal dynamics for this case are also shown in the accompanying video.}
\label{fig:Fig2_NonlinearProfile}
\end{figure}

In order to assess how a large-amplitude pressure wave divides across a symmetric junction, we first illustrate baseline examples using both the linear tube law (\emph{i.e.} $m=1$, $n=0$, Fig.~\ref{fig:Fig2_LinearProfile}) and the nonlinear tube law (\emph{i.e.} $m=10$, $n=3/2$, Fig.~\ref{fig:Fig2_NonlinearProfile}). In each case we plot several spatial profiles along both the parent and daughter vessels (Fig.~\ref{fig:Fig2_LinearProfile}a,\ref{fig:Fig2_NonlinearProfile}a) as well as a timetrace of the vessel cross-sectional area at selected points along the vessels (Fig.~\ref{fig:Fig2_LinearProfile}b,\ref{fig:Fig2_NonlinearProfile}b). An animation showing these dynamics is provided as online supplementary material.  In both cases the behaviour is qualitatively similar: the pressure increase at the inlet drives a rapid expansion of the entrance to the parent vessel (see timetraces in Fig.~\ref{fig:Fig2_LinearProfile}b, \ref{fig:Fig2_NonlinearProfile}b blue solid line), triggering a pressure wave which propagates along the parent vessel towards the junction (Fig.~\ref{fig:Fig2_LinearProfile}a, Fig.~\ref{fig:Fig2_NonlinearProfile}a). This pressure wave steepens as it propagates, forming a nonlinear shock wave (or elastic jump): as the wave propagates the cross-sectional area abruptly increases at each spatial location (see time-traces in Fig.~\ref{fig:Fig2_LinearProfile}b, \ref{fig:Fig2_NonlinearProfile}b, red solid line). Behind the elastic jump the vessel amplitude remains approximately constant, equal to that set by the driving pressure. Upon encountering the junction, part of the propagating pressure wave is transmitted into the two (identical) daughter vessels and part is reflected back along the parent towards the inlet (Fig.~\ref{fig:Fig2_LinearProfile}a, Fig. \ref{fig:Fig2_NonlinearProfile}a red dashed line). This reflection is associated with a large increase in flow through the junction (Fig.~\ref{fig:Fig2_LinearProfile}aii, Fig. \ref{fig:Fig2_NonlinearProfile}aii, red dashed line). The pressure wave transmitted into the two daughter vessels forms an elastic jump as it propagates, proceeding with almost constant amplitude towards the outlets (Fig.~\ref{fig:Fig2_LinearProfile}a, Fig. \ref{fig:Fig2_NonlinearProfile}a, green dotted line). However, the reflected wave in the parent vessel takes the form of a rarefaction wave which propagates slowly back toward the system inlet. In this example the corresponding local Mach number of the flow is largest close to the junction but the flow always remains subcritical (Fig.~\ref{fig:Fig2_LinearProfile}aiii, Fig. \ref{fig:Fig2_NonlinearProfile}aiii). In summary, in the examples the pressure perturbation at the inlet triggers a propagating elastic jump which is both reflected and transmitted at the junction, forming two new propagating elastic jumps in the daughter vessels and a reflected rarefaction wave in the parent vessel. The qualitative behaviour is independent of the choice of tube law. 

\begin{figure}
\centering
\includegraphics[width=6in]{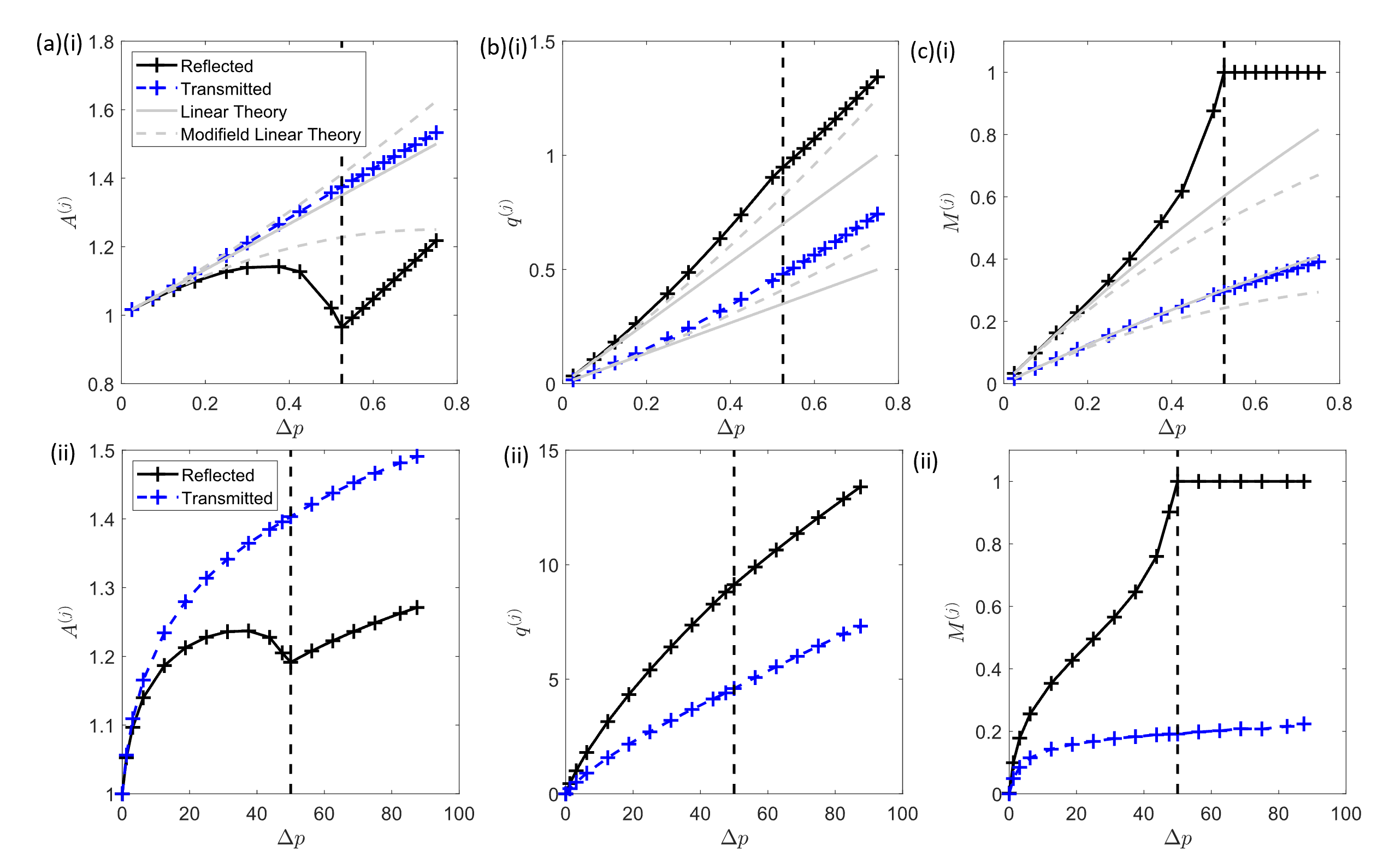}
\caption{Predictions of the flow on either side of the junction as a function of dimensionless driving pressure using the (i) linear tube law ($m=1$, $n=0$) and (ii) the nonlinear tube law ($m=10$, $n=3/2$). Predictions of the fully nonlinear model show the amplitude of the transmitted wave (blue $+$, with points joined by a dashed line) and the reflected wave (black $+$, with points joined by a solid line) in terms of: (a) vessel cross-sectional area; (b) fluid flux and (c) Mach number. Predictions of the first-order (grey solid line) and second-order linear (grey dotted line) theories are shown for the linear tube law in (i). The dotted line in each panel indicates the point where the flow in the parent vessel becomes transcritical, \emph{i.e.} $\Delta p=\Delta p_c$.} 
\label{fig:BifurcationDiagram_NonLinear}
\end{figure}

In order to quantify the transmission and reflection as the pressure wave encounters the junction, Fig.~\ref{fig:BifurcationDiagram_NonLinear} illustrates the amplitudes of the reflected and transmitted waves as a function of the driving pressure. As before, the behaviour is qualitatively similar for the linear tube law (Fig.~\ref{fig:BifurcationDiagram_NonLinear}i panels) and the nonlinear tube law (Fig.~\ref{fig:BifurcationDiagram_NonLinear}ii panels). As the driving pressure increases from zero, the amplitude of the elastic jump propagating along the parent increases accordingly, with a corresponding increase in the reflected and transmitted waves at the junction, evident in both the vessel cross-sectional areas (Fig.~\ref{fig:BifurcationDiagram_NonLinear}a) and the fluid flux through the junction (Fig.~\ref{fig:BifurcationDiagram_NonLinear}b), alongside an increase in the local Mach number around the junction (Fig.~\ref{fig:BifurcationDiagram_NonLinear}c). However as the input driving pressure increases further, the reflected cross-sectional area reaches a maximum and then starts to decline (Fig.~\ref{fig:BifurcationDiagram_NonLinear}a), despite a continued increase in the other amplitudes (Fig.~\ref{fig:BifurcationDiagram_NonLinear}a-c). In this regime the reflected Mach number in the parent vessel close to the junction increases rapidly for only small changes in the driving pressure, with the Mach number in the parent approaching unity for a finite value of the driving pressure (Fig. \ref{fig:BifurcationDiagram_NonLinear}c); we denote this limit point as $\Delta p_c$. Beyond this critical driving pressure (\emph{i.e.} $\Delta p > \Delta p_c$) there is a noticeable change in behaviour. The local Mach number at the junction in the parent remains fixed at $M_p^{(j)}=1$ at the boundary between subcritical and supercritical (Fig.~\ref{fig:BifurcationDiagram_NonLinear}c), the reflected rarefaction in the parent vessel remains pinned to the junction and reflected cross-sectional area in the parent increases almost linearly with the driving pressure (Fig.~\ref{fig:BifurcationDiagram_NonLinear}a). Interestingly, flow in the daughter vessels shows no indication of this limit point behaviour, where both the flow velocity and vessel cross-sectional area at the junction continue to increase with the driving pressure (Fig.~\ref{fig:BifurcationDiagram_NonLinear}). In summary, as the driving pressure increases the flow across the junction can eventually become transcritical, beyond which the reflected rarefaction wave in the parent vessel becomes pinned to the junction.

Numerical issues limit our ability to push this numerical method to larger driving pressures than those considered here. While the numerical method is stable for low (subcritical) driving pressures (\emph{i.e.} $\Delta p <\Delta p_c$), beyond the critical point $\Delta p_c$ the numerical method exhibits time-dependent oscillations and/or extremely sharp gradients in the spatial profiles at the mesh points around the junction, consistent with previous observations for supercritical flows \cite{Zhang2017}.

In order to validate these numerical observations, in Fig.~\ref{fig:BifurcationDiagram_NonLinear}(i panels) we compare our predictions with the linear tube law to those from the second-order analytical theory (Sec.~\ref{sec:model}\ref{sec:lin}, Appendix \ref{App_PertTheory}). As expected, we observe good agreement between the analytical and numerical approaches for low driving pressures. However, predictions of the two approaches diverge from one another well before the limit point. Our modified linear theory significantly improves the prediction of the reflected cross-sectional area at the junction for intermediate driving pressures (Fig.~\ref{fig:BifurcationDiagram_NonLinear}ai), at the expense of a slight reduction in agreement in other metrics (Fig.~\ref{fig:BifurcationDiagram_NonLinear}ci). However, this weakly nonlinear theory shows no evidence of transcriticality, and so to explain this transition we now proceed to derive a formal Riemann problem at the junction, analogous to the junction Riemann problems considered by others \cite{contarino2016junction,Murillo2021}. This Riemann problem is derived in Sec.~\ref{sec:riemann}, and results are presented alongside a comparison to the predictions of the finite volume method.

\section{Riemann problem at a symmetric junction}
\label{sec:riemann}

We construct the full Riemann problem in the neighbourhood of the junction, with the setup illustrated in Fig \ref{fig:Riemann_Profiles}(a). In line with classical Riemann problems in flexible-walled vessels (\emph{e.g.} \cite{Brook1999,toro2012simplified}) we use the same nonlinear governing equations (\ref{eq:GovEqu_Mass}-\ref{eq:GovEqu_Pressure}) but now prescribe a piece-wise uniform initial profile of both the cross-sectional area of the vessel and the fluid velocity (denoted by capital caligraphic letters), but with an imposed discontinuity in one or both of these quantities coincident with the junction ($x=0$). Hence, we apply the initial condition
\begin{align}
A_p(x,0)&={\cal A}_p, & u_p(x,0)&={\cal U}_p, & &(x<0),\\
A_d(x,0)&={\cal A}_d, & u_d(x,0)&={\cal U}_d, & &(x>0).
\end{align}
Any disturbance will propagate outwards from the point of initial discontinuity, and at any later time $t>0$ these initial conditions must still hold far enough away from the junction; these far-field boundary conditions replace the inlet and outlet boundary conditions considered in Sec.~\ref{sec:results}. Hence, upstream pressure and flow conditions are imposed through changes in the far-field conditions ${\cal A}_p$ and ${\cal U}_p$. Note that the corresponding initial wavespeeds ${\cal C}_i$ ($i=p,d$) can be calculated from Eq.~(\ref{Elastic_Wavespeed}).

As time progresses this initial discontinuity breaks up into left- and right--propagating waves with an intermediate region (often known as the `star' region in shallow water problems) surrounding the junction where we must solve for the (spatially uniform) vessel cross-sectional area and fluid velocity. The division of fluid (and any jump in material properties) across the junction means that only two quantities can be conserved, which we choose to be the mass and total pressure (in line with boundary conditions (\ref{Eq_Junc_Mass},\ref{Eq_Junc_TotPres})). Hence, this intermediate region must have a stationary wave at the junction ($x=0$) to accommodate the jump in the other variables (this stationary wave is essentially a zero eigenvalue of the system when formulated as a matrix problem \cite{toro2012simplified}). When expressed in terms of the model variables, we denote the piecewise uniform solution across this intermediate region as
\begin{align}
A_p&={\cal A}_{p}^{(j)}, & u_p&={\cal U}_{p}^{(j)}, & &(x_p(t)<x<0),\\
A_d&={\cal A}_{d}^{(j)}, & u_d&={\cal U}_{d}^{(j)}, & &(0<x<x_d(t)),
\end{align}
where $x_p(t)$ and $x_d(t)$ denote the edges of this intermediate region in the parent and daughter vessels, respectively; these edges are computed as part of the analysis below. The corresponding wavespeeds across the intermediate region, ${\cal C}^{(j)}_i$ ($i=p,d$) can be calculated from Eq.~(\ref{Elastic_Wavespeed}).

Using the notation of the Riemann problem, our boundary conditions of mass and total pressure conservation (\ref{Eq_Junc_Mass}-\ref{Eq_Junc_TotPres}) take the form
\begin{align}
& r_p {\cal A}_{p}^{(j)} {\cal U}_{p}^{(j)} = r_d {\cal A}_{d}^{(j)} {\cal U}_{d}^{(j)}, \notag  \\
  & \tfrac{1}{2}{\cal U}^{(j)^2}_{p} + \Gamma({\cal A}_{p}^{(j)}) = \tfrac{1}{2}{\cal U}_{d}^{(j)^2} + k_{d} \Gamma({\cal A}_{d}^{(j)}), \label{eq:junctionriemann}
\end{align}
where $r_p=1$ and $r_d=2$, as before.

Linking this intermediate region to the two far-field regions (where the flow takes the initial values) are left- and right-propagating waves. In a classical Riemann problem these are either a rarefaction wave or a shock wave, leading to four possible responses when the bifurcation is symmetric (Sec.~4\ref{Sec_RiemannStandardCases}), each a generalisation of the four standard responses in a single blood vessel \cite{Brook1999,toro2012simplified}. However, for flow in a flexible-walled vessel with an abrupt discontinuity in elastic stiffness, additional (resonant) flow structures are possible, as discussed by Han {\it et al.} (2015) \cite{han2015subcritical,han2015supercritical}. Two of these resonant cases prove relevant to the analysis of our symmetric junction system presented below, and emerge when the rarefaction wave in the parent vessel becomes transcritical, filling the entire space between the upstream far-field region and the stationary wave at the junction; this resonant interaction leads to at least one additional shock wave in the system. More details on these resonant cases are given in Secs.~\ref{sec:riemann}\ref{Sec:PR0DS} and \ref{sec:riemann}\ref{sec_PRDSS} below. A more complete study of the parameter space, involving several more families of resonances, has recently been conducted \cite{onah2023thesis}.

We now discuss the the governing equations relevant to each case in turn.  

\subsection{Governing equations for shocks and rarefactions}

The full governing equations (\ref{eq:GovEqu_Mass}-\ref{eq:GovEqu_Momentum}) can be expressed in Riemann invariant form \cite{Pedleybook}
\begin{align}
&\left(\frac{\partial}{\partial t}+(u_i\pm c_i)\frac{\partial}{\partial x}\right)R_{i\pm} =0, & R_{i\pm}=u_i\pm\int_{\overline{A}_i}^{A_i} \frac{c_i(A_i^*)}{A_i^*}\,\text{d} A_i^*,  
\label{Riemann_Invariant}
\end{align}
where $c_i$ is the nonlinear wavespeed defined in Eq.~(\ref{Elastic_Wavespeed}) and $R_{i\pm}$ are the corresponding Riemann invariants of the flow ($i=p,d$). By definition, these Riemann invariants remain unchanged along the characteristic curves in the $(x,t)$ plane governed by
\begin{equation}\label{H1f}
\frac{dx}{dt}=u_i \pm c_i, \qquad (i=p,d).
\end{equation}

The discontinuous initial profile breaks up into constituent waves, which can either be a shock wave or a rarefaction, propagating into either the parent vessel or the daughter vessels. We now consider shock waves and rarefaction waves in turn.

For a shock wave propagating steadily along vessel $i$  with velocity $\dot{s}_i$ (separating the far-field region and the intermediate region), conservation of mass and conservation of momentum across the shock front \cite{whitham,Pedleybook} results in the constraints
\begin{align}
\label{H1h}
\dot{s}_i= & \pm\frac{\mathcal{A}_{i}^{(j)}\mathcal{U}_{i}^{(j)}-\mathcal{A}_{i}\mathcal{U}_i}{\mathcal{A}_{i}^{(j)}-\mathcal{A}_i} 
\\ = &
\pm \frac{\mathcal{A}_{i}^{(j)}\mathcal{U}_{i}^{(j)^2}-\mathcal{A}_i\mathcal{U}_i^2
+\frac{m}{m+1}\left(\mathcal{A}_{i}^{(j)^{m+1}} - \mathcal{A}_i^{m+1}\right) - \frac{n}{n-1}\left(\mathcal{A}_{i}^{(j)^{-n+1}} - \mathcal{A}_i^{-n+1}\right)}{\mathcal{A}_{i}^{(j)}\mathcal{U}_{i}^{(j)}-\mathcal{A}_i\mathcal{U}_i},
\end{align}
for $i=p,d$. The requirement for the shock to be propagating away from the junction means that for the parent vessel $(i=p)$ the shock speed must be negative while for the daughter vessels ($i=d$) the shock speed must be positive. However, in both cases these two conditions combine to the same nonlinear constraint linking the far-field conditions to the flow in the intermediate region in the form
\begin{equation}
(\mathcal{U}_{i}^{(j)}-\mathcal{U}_i)^2=\left[\frac{m}{m+1}\left(\mathcal{A}_{i}^{(j)^{m+1}}-\mathcal{A}_i^{m+1}\right)+ \frac{n}{n-1}\left(\mathcal{A}_i^{-n+1}-\mathcal{A}_{i}^{(j)^{-n+1}}\right)\right]\frac{\left(\mathcal{A}_{i}^{(j)}-\mathcal{A}_i\right)}{\mathcal{A}_i\mathcal{A}_{i}^{(j)}}, \label{GeneralShock}
\end{equation}
for $i=p,d$. To be a physical solution this shock wave must satisfy the appropriate Lax entropy conditions \cite{lax1972formation}, which for the parent vessel take the form
\begin{align}
 \left(\mathcal{U}^{(j)}_p-\mathcal{C}^{(j)}_p\right) \le \dot{s}_p \le (\mathcal{U}_p-\mathcal{C}_p),  \label{eq:LaxConditionsp} 
\end{align}
and for the daughter vessels take the form
\begin{align}
(\mathcal{U}_d+\mathcal{C}_d) \le \dot{s}_d \le \left(\mathcal{U}^{(j)}_d+\mathcal{C}^{(j)}_d\right). \label{eq:LaxConditionsd}
\end{align}

In contrast, for a rarefacton wave linking the far-field region in the parent vessel to the junction region, the spatial profiles of the wave vary over a spatial interval. Any rarefaction adjacent to the far-field must be bounded by straight-line characteristics emanating from the origin of the $x-t$ plane, and so a rarefaction in the parent vessel is confined to the interval $\left({\cal U}_{p}-{\cal C}_{p}\right)t <x< \left({\cal U}^{(j)}_{p}-{\cal C}^{(j)}_p\right)t$, while any rarefaction in the daughter vessels is confined to the interval $\left({\cal U}_{d}+{\cal C}_{d}\right)t < x < \left({\cal U}^{(j)}_{d}+{\cal C}^{(j)}_d\right)t$. To fix the uniform values of the flow variables across the intermediate regions (on the junction side of the rarefaction) we follow along characteristics leaving the far-field region into the rarefaction toward the junction. The argument is standard \cite{whitham,onah2023thesis}, and it emerges that we can compute implicit equations for the spatial profiles across a rarefaction adjacent to the far-field in either the parent vessel or the daughter vessels. For example, for every $x$ and $t$ within the rarefaction fan in the parent vessel we can express the cross-sectional area profile $A^{(r)}_{p}(x,t)$ as the solution of the implicit equation derived from the negative Riemann invariant
\begin{align}
\label{eq:profilerarep}
\frac{x}{t} + c_p\left(A^{(r)}_{p}\right) + \int_{\mathcal{A}_p}^{A^{(r)}_p} \frac{c_p(A^\ast)}{A^\ast} \,\text{d}A^\ast = \mathcal{U}_p.
\end{align}
Similarly, for every $x$ and $t$ within the rarefaction fan in the daughter vessel we can express the cross-sectional area profile $A^{(r)}_{d}(x,t)$ as the solution of the implicit equation derived from the positive Riemann invariant
\begin{align}
\frac{x}{t} -c_d\left(A^{(r)}_{d}\right) - \int_{\mathcal{A}_d}^{A^{(r)}_d} \frac{c_d(A^\ast)}{A^\ast} \,\text{d}A^\ast = \mathcal{U}_d.
\end{align}
Simultaneously, we derive nonlinear constraints linking the far-field conditions to the intermediate variables by imposing continuity along the edge of the rarefaction meeting the intermediate region. These take the form
\begin{align}
{\cal U}^{(j)}_{p} &= {\cal U}_p -  \int_{{\cal A}_p}^{{\cal A}^{(j)}_{p}} \frac{c_p(A^\ast)}{A^*}\,\text{d}A^*,
\label{GeneralRarefactionp}\\
{\cal U}^{(j)}_{d} &= {\cal U}_d + \int_{{\cal A}_d}^{{\cal A}^{(j)}_{d}} \frac{c_d(A^\ast)}{A^*}\,\text{d}A^*.
\label{GeneralRarefactiond}
\end{align}
These rarefaction waves cannot cross the stationary wave at the junction and must not outrun the far-field solution, and so for the solution to remain physical we impose the constraints
\begin{align}
{\cal U}_{p} - {\cal C}_{p} &< {\cal U}_{p}^{(j)} - {\cal C}_{p}^{(j)} <0,  \label{eq:RarefactionPhysicalp} \\
0& < {\cal U}_{d}^{(j)} + {\cal C}_{d}^{(j)} < {\cal U}_{d} + {\cal C}_{d}. \label{eq:RarefactionPhysicald}
\end{align}
The special case where the parent rarefaction becomes transcritical at the junction (\emph{i.e.} ${\cal U}_{p}^{(j)}={\cal C}_{p}^{(j)}$) is considered in Sec.~\ref{sec:riemann}\ref{Sec:PR0DS} and Sec.~\ref{sec:riemann}\ref{sec_PRDSS} below.

\subsection{Four standard responses consisting of single shocks and rarefactions} 
\label{Sec_RiemannStandardCases}

The standard responses in a classical Riemann problem consist of a single shock or rarefaction in each of the parent and daughter vessels. Allowing for every combination produces four possible cases. In these four standard examples we solve for four unknowns: ${\cal A}_{p}^{(j)}$,  ${\cal A}_{d}^{(j)}$, ${\cal U}_{p}^{(j)}$ and ${\cal U}_{d}^{(j)}$ based on the four far-field conditions (${\cal A}_p$, ${\cal U}_p$ ${\cal A}_d$ and ${\cal U}_d$). The two flow velocities at the junction can be expressed in terms of the corresponding cross-sectional areas using either Eq.~(\ref{GeneralShock}) for a shock wave or Eq.~(\ref{GeneralRarefactionp},\ref{GeneralRarefactiond}) for a rarefaction wave. The junction cross-sectional areas are then calculated numerically using Newton's method by imposing the junction boundary conditions (\ref{eq:junctionriemann}). Our strategy is to consider which solutions are physically meaningful by imposing Lax conditions (\ref{eq:LaxConditionsp},\ref{eq:LaxConditionsd}) on shock waves and physicality conditions (\ref{eq:RarefactionPhysicalp},\ref{eq:RarefactionPhysicald}) on rarefaction waves.

\begin{figure}
\centering
\includegraphics[width=6in]{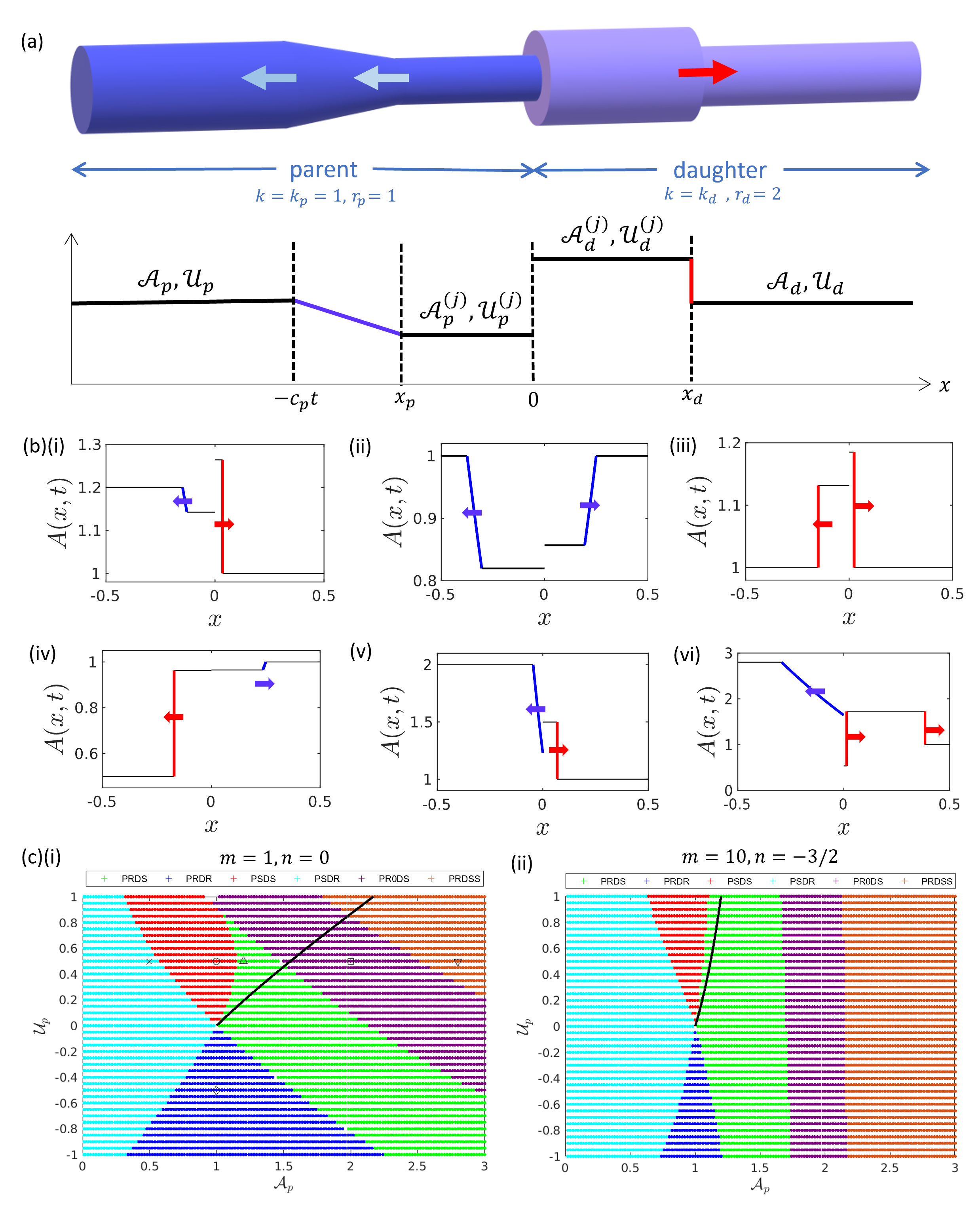}
\caption{Junction Riemann problem. (a) Notation for the junction Riemann problem (illustrated for a case with a rarefaction in the parent vessel and shock in the daughter vessels); (b) typical spatial profiles for the six different solutions to the junction Riemann problem (assuming a linear tube law, $m=1$, $n=0$) at the six points shown parameter space (c,i) below at time $t=0.25$: (i) PRDS (triangle in (c,i)); (ii) PRDR (diamond in (c,i)); (iii) PSDS (circle in (c,i)); (iv) PSDR (cross in (c,i)); (v) PR0DS (square in (c,i)); (vi) PRDSS (inverted triangle in (c,i)); (c) Summary of the junction Riemann problem across the parameter space spanned by the far-field conditions in the parent vessel (${\cal A}_p$, ${\cal U}_p$) holding the far-field conditions in the daughter vessel fixed (${\cal A}_d=1$, ${\cal U}_d=0$) for the (i) linear tube law ($m=1$, $n=0$) and (ii) nonlinear tube law ($m=10$, $n=-3/2$), where the colour of the circle indicates the form of the response. In each panel in (b) the rarefaction waves are shown in blue and the shock waves are shown in red.} 
\label{fig:Riemann_Profiles}
\end{figure}

Examples of the four standard cases are plotted in Fig.~\ref{fig:Riemann_Profiles}(b): rarefaction in the parent vessel and a shock in each daughter vessel (PRDS, Fig.~\ref{fig:Riemann_Profiles}bi); rarefaction in the parent vessel and a rarefaction in each daughter vessel (PRDR, Fig.~\ref{fig:Riemann_Profiles}bii) shock in the parent vessel and a shock in each daughter vessel (PSDS, Fig.~\ref{fig:Riemann_Profiles}biii); shock in the parent vessel and a rarefaction in each daughter vessel (PSDR, Fig.~\ref{fig:Riemann_Profiles}biv). The points in parameter space corresponding to these four examples are highlighted on Fig.~\ref{fig:Riemann_Profiles}(c). It emerges below that these four classical cases are insufficient to span the parameter space. Two further cases, where resonance between a rarefaction wave in the parent vessel and the stationary wave at the junction produces an additional shock wave are discussed below, where this new shock wave can be of zero speed pinned to the junction (Sec.~\ref{Sec:PR0DS}) or propagating into the daughter vessels (Sec.~\ref{sec_PRDSS}).

\subsection{Resonance with a zero speed shock: PR0DS}
\label{Sec:PR0DS}

For the parameter values considered below, in PRDS the rarefaction wave in the parent vessel case can expand all the way to the junction and interact with the stationary wave, producing an additional zero speed shock wave at the junction; we denote this case PR0DS. This is analogous to Class C resonance reported by Han \emph{et al.} (2015) for a vessel with an abrupt discontinuity in elastic stiffness \cite{han2015subcritical}. Here, the stationary wave at $x=0$ splits into two stationary waves an asymptotically small distance $\epsilon \ll 1$ up and downstream of the junction, located at $x=\pm \epsilon$ respectively. This division then introduces an additional resonant region around the junction. Following Han \emph{et al.} (2015) \cite{han2015subcritical}, we expect that the stationary wave just inside the parent (at $x=-\epsilon$) will be supercritical (\emph{i.e.} the flows on either side will be supercritical) while the stationary wave just inside the daughter vessels (at $x=\epsilon$) will be the subcritical (\emph{i.e.} the flows on either side will be subcritical). 

In the asymptotically small inner region between the two stationary waves  ($-\epsilon \le x \le \epsilon$) we denote the flow variables with the superscript $^{(s)}$. Across this region the material and environmental properties are spatially uniform, but take values intermediate to those in the parent and daughter vessels. In this study we allow for discontinuities in the vessel stiffness, where the intermediate value is denoted $k_s$, and the mass distribution parameter across the junction, where the intermediate value is denoted $r_s$. The vessel baseline cross-sectional area and external pressure can vary similarly, but here they are assumed equal in the daughter and parent vessels. We assume that these unknown properties between the stationary waves (\emph{i.e.} those with a subscript $s$) can be scaled linearly between their values in the parent and daughter vessels using a single interpolation parameter $s$, such that $s=0$ denotes the properties of the parent vessel and $s=1$ denotes the properties of the daughter vessels; these intermediate values can be written
\begin{align} 
k_s =& k_p + s(k_d-k_p), \qquad r_s = r_p + s(r_d-r_p),  \qquad (0\le s\le 1).
\end{align}

In the resonant state, we immediately determine the flow variables on the parent side of the supercritical stationary wave by taking the outer limit ($x\rightarrow \epsilon^-$) of the rarefaction profiles across the parent (\emph{e.g.} Eq.~(\ref{eq:profilerarep})), where the critical value of the parent cross-sectional area ${\cal A}^{(j)}_p$ satisfies the implicit equation
\begin{align}
\label{eq:critAres}
c_p\left({\cal A}^{(j)}_{p}\right) + \int_{\mathcal{A}_p}^{{\cal A}^{(j)}_p} \frac{c_p(A^\ast)}{A^\ast} \,\text{d}A^\ast = \mathcal{U}_p,
\end{align}
with error $O(\epsilon)$, and it follows that
\begin{align}
\label{eq:critucres}
{\cal U}_{p}^{(j)} &= {\cal C}_{p}^{(j)} = \left( m{\cal A}_{p}^{(j)^m} + n{\cal A}_{p}^{{(j)}^{-n}} \right)^{1/2} +O(\epsilon). 
\end{align}

Across the supercritical stationary wave ($x=-\epsilon$) we impose conservation of mass and momentum conditions (analogous to Eq.~\ref{eq:junctionriemann})
\begin{equation}
r_p {\cal A}_{p}^{(j)} {\cal U}_{p}^{(j)}= r_s {\cal A}_{p}^{(s)} {\cal U}_{p}^{(s)},  \qquad \tfrac{1}{2}{\cal U}_{p}^{(j)^2} + k_p \Gamma({\cal A}_{p}^{(j)}) =  + \tfrac{1}{2}{\cal U}_{p}^{(s)^2} + k_s \Gamma({\cal A}_{p}^{(s)}).\label{eq:csupstatb}
\end{equation}
Across the zero-speed shock at $x=0$ we impose shock relations similar to (\ref{H1h}) in the form
\begin{align}
\label{eq:czero}
&\frac{\mathcal{A}_{p}^{(s)}\mathcal{U}_{p}^{(s)}-\mathcal{A}^{(s)}_{d}\mathcal{U}^{(s)}_d}{\mathcal{A}_{p}^{(s)}-\mathcal{A}^{(s)}_d} =  \nonumber \\
& \frac{\mathcal{A}_{p}^{(s)}(\mathcal{U}_{p}^{(s))^2}-\mathcal{A}^{(s)}_d(\mathcal{U}^{(s)}_d)^2 +\frac{m}{m+1}\left((\mathcal{A}_p^{(s)})^{m+1} - (\mathcal{A}^{(s)}_d)^{m+1}\right) - \frac{n}{n-1}\left((\mathcal{A}^{(s)}_p)^{-n+1} - (\mathcal{A}^{(s)}_d)^{-n+1}\right)}{\mathcal{A}_{p}^{(s)}\mathcal{U}_{p}^{(s)}-\mathcal{A}_d^{(s)}\mathcal{U}_d^{(s)}}.
\end{align}
Similarly, we impose conservation of mass and momentum conditions (analogous to Eq.~(\ref{eq:junctionriemann})) across the subcritical stationary wave ($x=\epsilon$) in the form
\begin{equation}
r_d {\cal A}_{d}^{(j)} {\cal U}_{d}^{(j)}= r_s {\cal A}_{d}^{(s)} {\cal U}_{d}^{(s)},  \qquad \tfrac{1}{2}{\cal U}_{d}^{(j)^2} + k_d \Gamma({\cal A}_{d}^{(j)})=  + \tfrac{1}{2}{\cal U}_{d}^{(s)^2} + k_s \Gamma({\cal A}_{d}^{(s)}). \label{eq:csubstatb}
\end{equation}

Hence, to fully determine the PR0DS resonant profile we use Newton's method to solve the jump conditions across the supercritical stationary wave (Eq.~\ref{eq:csupstatb}), the nonlinear constraint across the zero speed shock (Eq.~\ref{eq:czero}) and the jump conditions across the subcritical stationary wave (Eq.~\ref{eq:csubstatb}) coupled to the condition for a shock in the daughter vessels (Eq.~(\ref{GeneralShock}) with $i=d$) for the six unknowns ${\cal U}_{p}^{(s)}$, ${\cal A}_{p}^{(s)}$, ${\cal U}_{d}^{(s)}$, ${\cal A}_{d}^{(s)}$ and ${\cal A}_{d}^{(j)}$ and ${\cal U}_{d}^{(j)}$.

This resonant case is subject to physicality conditions on the transcritical rarefaction in the form,
\begin{equation}
{\cal U}_{p} - {\cal C}_{p} < {\cal U}_{p}^{(j)} - {\cal C}_{p}^{(j)},
\end{equation}
in addition to Lax conditions on the right-propagating shock in the daughter vessel (Eq.~(\ref{eq:LaxConditionsd})).

A typical spatial profile for an example of PR0DS resonance is presented in Fig.~\ref{fig:Riemann_Profiles}(a,v), where the asymptotic region between the two stationary waves is invisible. The point in parameter space corresponding to this example is highlighted on Fig.~\ref{fig:Riemann_Profiles}(c). 

\subsection{Resonant flow with a new shock in each daughter vessel: PRDSS} 
\label{sec_PRDSS}

Well beyond criticality, the zero speed shock in the PR0DS case (see Sec.~\ref{sec:riemann}\ref{Sec:PR0DS}) becomes mobile, splitting across the junction into the two daughter vessels as it propagates. This case is denoted PRDSS and is analogous to the Class D resonance discussed by Han {\it et al.} for a single vessel with discontinuous elastic stiffness \cite{han2015subcritical,han2015supercritical,onah2023thesis}. Here, there is a single stationary wave at $x=0$, and the outer limit of flow variables across the transcritical rarefaction again satisfies Eq.~(\ref{eq:critAres},\ref{eq:critucres}) with $\epsilon=0$. Denoting the position of the new shock front in the daughter vessels as $x=s_a(t)$, we now sub-divide the intermediate region in the daughter vessel into two (spatially uniform) pieces. Across the region between the stationary wave at $x=0$ and the resonant shock ($x=s_a$) we denote the (spatially uniform) flow variables with the superscript $^{(j)}$ (in accordance with our naming convention above), while in the region between this new shock wave ($x=s_a$) and the existing shock wave ($x=s_d$) we denote the (spatially uniform) flow variables with the superscript $^{(a)}$. 

The conservation conditions across the stationary wave at the junction (\ref{eq:junctionriemann}) take the form
\begin{align}
\label{eq:dstatb}
r_p {\cal A}_{p}^{(j)} {\cal U}_{p}^{(j)} &= r_d {\cal A}_{d}^{(j)} {\cal U}_{d}^{(j)},  \qquad \tfrac{1}{2}{\cal U}_{p}^{(j)^2} + \Gamma({\cal A}_{p}^{(j)}) = \tfrac{1}{2}{\cal U}_{d}^{(j)^2} + k \Gamma({\cal A}_{d}^{(j)}).
\end{align}
As before, applying conservation of mass and momentum across the front of the resonant shock in the daughter vessels gives two expressions for the propagation speed (similar to Eq.~\ref{H1h}) in the form 
\begin{align}
\label{eq:dshockspeed}
\dot{s}_a&=\frac{\mathcal{A}^{(j)}_{d}\mathcal{U}^{(j)}_{d}-\mathcal{A}^{(a)}_{d}\mathcal{U}^{(a)}_{d}}{\mathcal{A}^{(j)}_{d}-\mathcal{A}^{(a)}_{d}} =\nonumber\\
&
\frac{\mathcal{A}_d(\mathcal{U}^{(j)}_{d})^2 -\mathcal{A}^{(a)}_{d}(\mathcal{U}^{(j)}_{d})^2 +\frac{m}{m+1}\left((\mathcal{A}^{(j)}_{d})^{m+1}-(\mathcal{A}^{(a)}_{a})^{m+1}\right)-\frac{n}{n-1}\left((\mathcal{A}^{(j)}_{d})^{-n+1}-(\mathcal{A}^{(a)}_{d})^{-n+1}\right)}{\mathcal{A}^{(j)}_{d}\mathcal{U}^{(j)}_{d}-\mathcal{A}^{(a)}_{d}\mathcal{U}^{(a)}_{d}};
\end{align}
in a similar manner to Eq.~(\ref{GeneralShock}), these reduce to the single nonlinear constraint between the two pieces of the intermediate region in the daughter vessels
\begin{multline}
\label{eq:dnewshock}
(\mathcal{U}_{d}^{(j)}-\mathcal{U}_{d}^{(a)})^{^2}= \\ \left(\frac{m}{m+1}\left(\mathcal{A}_{d}^{(j)^{m+1}}-\mathcal{A}_{d}^{{(a)}^{m+1}}\right)- \frac{n}{n-1}\left(\mathcal{A}_{d}^{(j)^{-n+1}}-\mathcal{A}_{d}^{(a)^{-n+1}}\right)\right)\frac{\left(\mathcal{A}_{d}^{(j)}-\mathcal{A}_{d}^{(a)}\right)}{\mathcal{A}_{d}^{(j)}\mathcal{A}_{d}^{(a)}}.
\end{multline}

To fully determine the PRDSS resonant profile we use Newton's method to solve the two conditions across the stationary wave (Eq.~\ref{eq:dstatb}) and the constraint across the resonant shock (Eq.~\ref{eq:dnewshock}) coupled to the condition for the shock in the daughter vessel (Eq.~\ref{GeneralShock} with $i=d$) for the four unknowns ${\cal A}_{d}^{(j)}$, ${\cal U}_{d}^{(j)}$, ${\cal A}_{d}^{(a)}$ and ${\cal U}_{d}^{(a)}$. 

The pre-existing shock in the daughter vessels is subject to Lax conditions (Eq.~\ref{eq:LaxConditionsd}), whereas the new (resonant) shock is subject to Lax conditions based on the properties of the parent vessel (as this shock has originated from a resonance in the parent vessel) in the form
\begin{equation}
{\cal U}_d^{(a)} - {\cal C}_d^{(a)} \le \dot{s}_a \le {\cal U}_d^{(j)} - {\cal C}_d^{(j)}.
\end{equation}

A typical spatial profile for an example of PRDSS resonance is presented in Fig.~\ref{fig:Riemann_Profiles}(a,vi). The point in parameter space corresponding to this example is highlighted on Fig.~\ref{fig:Riemann_Profiles}(c).

\subsection{The junction Riemann problem across the parameter space}

To determine the solution of the Riemann problem at a given point in the parameter space spanned by the four far-field conditions (${\cal A}_p$, ${\cal U}_p$, ${\cal A}_d$, ${\cal U}_d$) our strategy is test for all six possible cases using a variety of initial conditions. We check each solution against their corresponding Lax and physicality conditions to determine if the solution is valid. Once a solution has been identified we use continuation methods to follow this solution branch until it terminates. In this way we construct large-scale sweeps of the parameter space and identify all feasible solutions. Across all cases tested we found no evidence of more than one solution at any point in the parameter space. However, our parameter searches were restricted to subcritical and transcritical flows, which are connected across the parameter space, and multi-valued solutions may become possible when the initial flow is strongly supercritical \cite{toro2012simplified,toro2013flow}.

In order to assess how these six different families of solutions for the junction Riemann problem co-exist, in Fig.~\ref{fig:Riemann_Profiles}(c) we plot an overview of the parameter space spanned by the far-field conditions in the parent vessel (\emph{i.e.} ${\cal A}_p$ and ${\cal U}_p$) assuming that far-field conditions in the daughter vessels are held fixed (${\cal A}_d=1$, ${\cal U}_d=0$); we use different coloured markers to indicate each family of solutions and compare the outcomes using the linear tube law ($m=1$, $n=0$ in Fig.~\ref{fig:Riemann_Profiles}c, left panel) to those from the more general nonlinear tube law ($m=10$, $n=3/2$ in Fig.~\ref{fig:Riemann_Profiles}c, right panel). In this first case we assume that all material and geometric properties of the constituent vessels are identical, and so the only discontinuity is due to the division of mass across the junction.  Around the point of undisturbed flow (\emph{i.e.} ${\cal A}_p=\overline{A}_p=1$ and ${\cal U}_p=0$) the system exhibits the four classical solutions in distinct (adjoining) wedges. Such a distribution of the classical solutions is also seen in the analysis of the Riemann problem in a single flexible-walled vessel \cite{onah2023thesis}. However, for larger values of the parent vessel far-field cross-sectional area, ${\cal A}_p$, the system adopts a PR0DS resonant solution. For the linear tube law the limiting value of ${\cal A}_p$ for the onset of resonance varies reduces almost linearly with increasing far-field velocity in the parent ${\cal U}_p$ (Fig.~\ref{fig:Riemann_Profiles}ci), while for the nonlinear tube law the limiting value of ${\cal A}_p$ is almost entirely independent of ${\cal U}_p$ (Fig.~\ref{fig:Riemann_Profiles}cii). Furthermore, for even larger values of the parent vessel far-field cross-sectional area, ${\cal A}_p$, the system transitions to PRDSS where in each case the limiting values of ${\cal A}_p$ follow a similar trend to the onset of transcriticality. In summary, the division of mass across the junction from the parent vessel into the daughter vessels is sufficient to induce resonance for sufficiently large values of the far-field cross-sectional area in the parent vessel.

It emerges that if, instead, the far-field conditions in the daughter vessels are varied and those in the parent vessel are held fixed, the resonant solutions occur for small values of ${\cal A}_d$ (not shown here, see \cite{onah2023thesis}), but this regime is not the focus of the present study.

\begin{figure}
\centering
\includegraphics[width=6.5in]{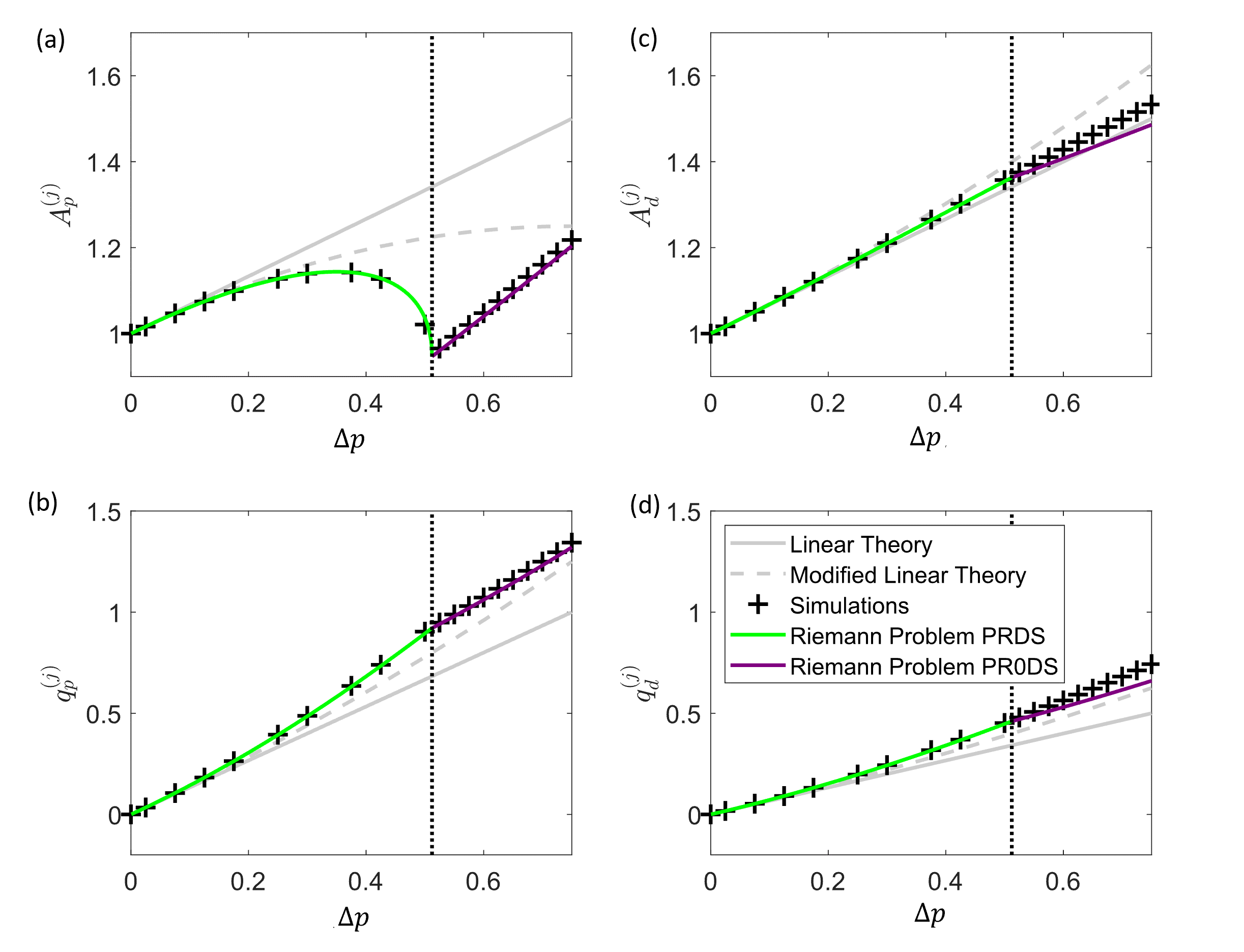}
\caption{Predictions of the flow properties either side of the junction for increasing driving pressure using multiple methods. The linear tube law ($m=1$, $n=0$) is used here. (a) cross-sectional area in the parent vessel; (b) flow rate in the parent vessel; (c) cross-sectional area in the daughter vessels; (d)  flow rate in the daughter vessel. In each panel the symbols are solutions from the full nonlinear computations, the solid lines are solutions of the junction Riemann problem where the line colour indicates the type of response and the grey solid (dashed) line is the prediction of the first-order (second-order) linear theory.}
\label{fig:CompareAnalytics}
\end{figure}

We note that the full numerical simulations in Sec.~\ref{sec:results} exhibit spatially uniform profiles in the far-field (both ahead of and behind the propagating waves, see examples in Figs.~\ref{fig:Fig2_LinearProfile}, \ref{fig:Fig2_NonlinearProfile}). This means that each simulation of the full numerical model effectively mimicks a Riemann problem at a single point in the parameter space spanned by the four far-field conditions (${\cal A}_p$, ${\cal U}_p$, ${\cal A}_d$, ${\cal U}_d$). For an initial shock wave propagating along the parent vessel driven by an inlet perturbation of amplitude $\Delta p$, the far-field conditions in the parent must satisfy the constraints (\ref{H1h}) separating the perturbed region behind and an undisturbed flow region ahead, so that
\begin{align}
&k_p \Gamma({\cal A}_p) = \Delta p, \qquad {\cal U}_p =\overline{u}_p +\frac{\left( \frac{m}{m+1}\left(1-{\cal A}_p^{m+1}\right) - \frac{n}{n-1}\left(1-{\cal A}_p^{-(n-1)}\right)\right)^{1/2} (1-{\cal A}_p)^{1/2} }{{\cal A}_p^{1/2}}.
\label{eq:shockline}
\end{align}
Hence, as the driving pressure increases the solution of Eq.~(\ref{eq:shockline}) traces a path through the parameter space spanned by the far-field conditions  (${\cal A}_p$, ${\cal U}_p$, ${\cal A}_d$, ${\cal U}_d$).  For example, for the bifurcation diagrams traced as a function of the driving pressure in Fig.~\ref{fig:CompareAnalytics}, the corresponding path through the parameter space of the Riemann problem is shown as a thick black line in Figs.~\ref{fig:Riemann_Profiles}(ci) and \ref{fig:Riemann_Profiles}(cii) for the linear and nonlinear tube laws, respectively.

In order to quantitatively compare the three approaches taken to study this system, Fig.~\ref{fig:CompareAnalytics} re-traces the bifurcation diagrams of Fig.~\ref{fig:BifurcationDiagram_NonLinear} with the linear tube law (the nonlinear tube law is qualitatively similar) and plots the predictions of the full numerical simulations against the corresponding predictions of the Riemann problem and the analytical theory, showing the change in the vessel cross-sectional areas at the junction (Fig.~\ref{fig:CompareAnalytics}a,c) and the flux through the junction (Fig.~\ref{fig:CompareAnalytics}b,d). As the driving pressure increases from zero the Riemann problem exhibits a PRDS solution (consistent with the spatial profiles of the full numerical simulations shown in Figs.~\ref{fig:Fig2_LinearProfile} and \ref{fig:Fig2_NonlinearProfile},) and the traces of the amplitude of the reflected and transmitted waves at the junction agree almost exactly with those from the full numerical simulations (Fig.~\ref{fig:CompareAnalytics}); as noted previously, the analytical theories agree well for small driving, but eventually diverge as the driving amplitude increases. Furthermore, the solution of the Riemann problem becomes transcritical at $\Delta p = \Delta p_c \approx 0.513$ (for the linear tube law) (Fig.~\ref{fig:CompareAnalytics}), almost exactly coincident with the predictions of the full numerical simulations, transitioning to a PR0DS solution. Beyond this limit point the predictions of the two approaches agree qualitatively, although there is an increasing discrepancy as $\Delta p$ increases (particularly in the flow through the daughter vessels, Fig.~\ref{fig:CompareAnalytics}b,d) which we attribute to the numerical issues encountered in the full problem caused by an inability to resolve the zero speed shock at the junction predicted by the Riemann problem. For even larger values of the driving pressure the Riemann problem exhibits PRDSS, but the full numerical solutions do not converge in this regime; it is not clear if this lack of convergence is due to an inability to resolve the additional shock waves in the daughter vessels, or because the larger driving pressure imposes much larger spatial gradients which are difficult to resolve. In summary, this figure demonstrates that the predictions of the full numerical model and the Riemann problem agree very well in the subcritical regime, but also how the predictions gradually deviate from each other beyond the onset of transcriticality; the characteristic extrapolation technique used to match across the junction in the full simulations develops numerical issues, and is not capable of predicting the additional shock wave(s) predicted by the Riemann problem.

We now proceed to consider changes in vessel stiffness on either side of the junction (Sec.~\ref{sec:stiffness}) and an underlying base flow through the network (Sec.~\ref{sec:baseflow}), comparing each with the original numerical model presented in Sec.~\ref{sec:results}. 


\section{A discontinuity in vessel stiffness across the junction}
\label{sec:stiffness}

\begin{figure}
\centering
\includegraphics[width=6.2in]{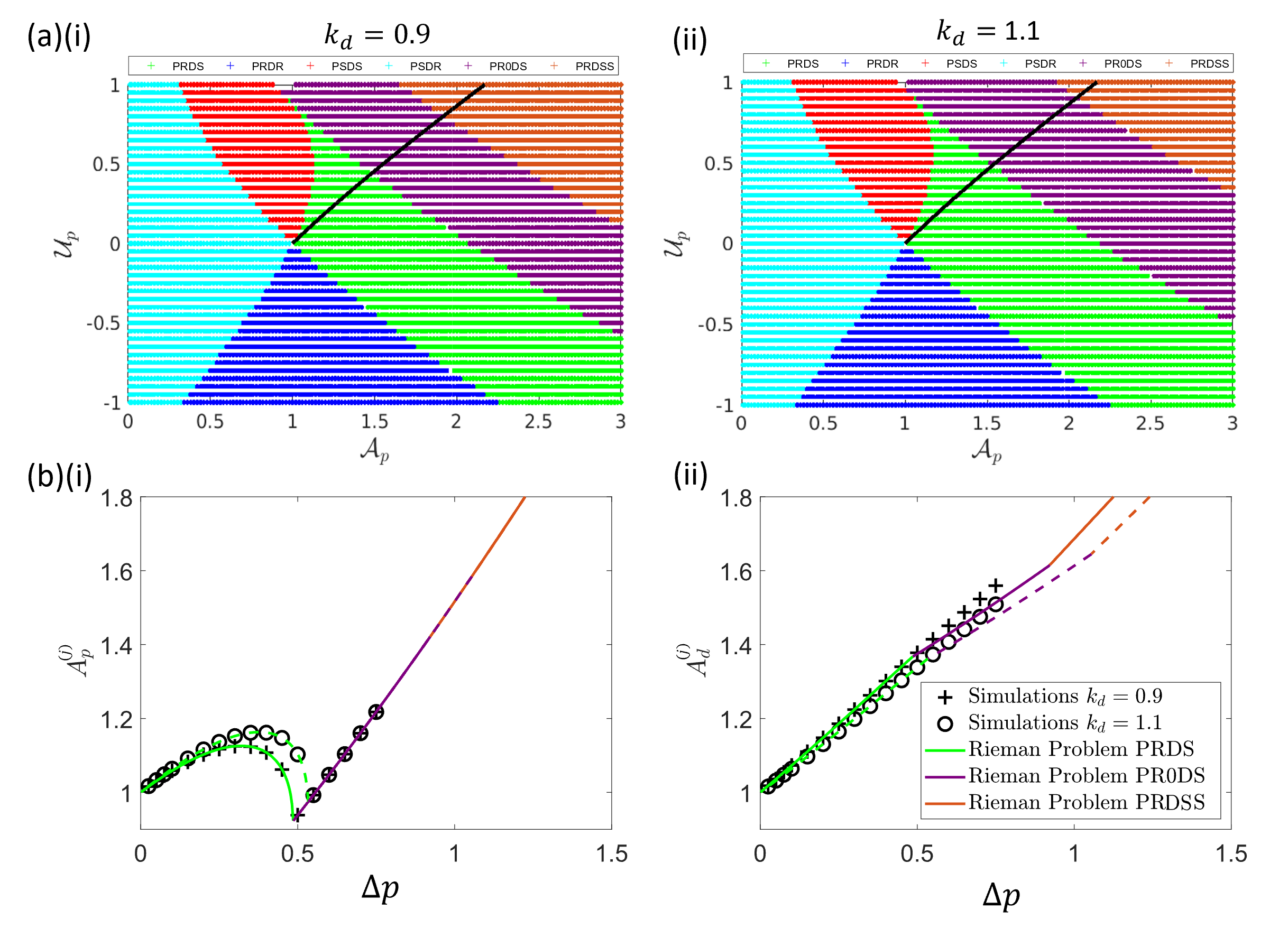}
\caption{Effect of an abrupt discontinuity in vessel stiffness coincident with the junction. The linear tube law ($m=1$, $n=0$) is used here. (a) solutions of the junction Riemann problem across the parameter space spanned by the far-field conditions in the parent vessel (${\cal A}_p$, ${\cal U}_p$) holding the far-field conditions in the daughter vessel fixed (${\cal A}_d=1$, ${\cal U}_d=0$) with (i) $k_d=0.9$ (ii) $k_d=1.1$; (b) tracing the properties of the flow on either side of the junction as a function of the dimensionless driving pressure comparing the full numerical simulations (symbols) with solutions for the junction Riemann problem for $k_d=0.9$ ($+$, solid lines) and $k_d=1.1$ ($o$ and dotted lines) showing: (i) cross-sectional area in the parent vessel and (ii) cross-sectional area in the daughter vessels. The black lines in (a) trace the path through the parameter space of the junction Riemann problem which mimic the full nonlinear simulations, Eq.~(\ref{eq:shockline}).}
\label{fig_stiffness}
\end{figure}

Blood vessel stiffness changes gradually through the circulatory network, where vessels generally become more compliant with increasing distance from the heart. In order to determine how a change in vessel stiffness influences the predictions of the model, and in particular the onset of transcriticality, Fig.~\ref{fig_stiffness} examines the behaviour of the system with the linear tube law (the nonlinear tube law is qualitatively similar) when the daughter vessels are less stiff than the parent (Fig.~\ref{fig_stiffness}i) and more stiff than the parent (Fig.~\ref{fig_stiffness}ii panels). In each case we plot the parameter space of the Riemann problem spanned by the far-field properties in the parent vessel (Fig.~\ref{fig_stiffness}a), isolate the path through the parameter space corresponding to the inlet perturbation from Eq.~(\ref{eq:shockline}) (black line in Fig.~\ref{fig_stiffness}a) and then compare the traces of the amplitude of the reflected wave in the parent and the transmitted wave in the daughters measured at the junction to those of the full numerical simulations (Fig.~\ref{fig_stiffness}b). The parameter spaces are qualitatively similar for the two different values of daughter vessel stiffness, but the onset of resonance is pushed to larger values of the parent cross-sectional area as the stiffness of the daughter vessels increases (Fig.~\ref{fig_stiffness}). A similar trend is evident in the bifurcation diagrams, where the limit point for the onset of both PR0DS and PRDSS are pushed to larger values of the driving pressure as the vessel stiffness increases (Fig.~\ref{fig_stiffness}b). Note that again we see increasing discrepancy between the full numerical simulations and the predictions of the Riemann problem once the driving pressure exceeds the limiting value (Fig.~\ref{fig_stiffness}b). In summary, having the daughter vessels more compliant than the parent vessel promotes the onset of resonance, occurring for lower values of the driving pressure.

\section{Base flow through the network}
\label{sec:baseflow}

\begin{figure}
\centering
\includegraphics[width=6.5in]{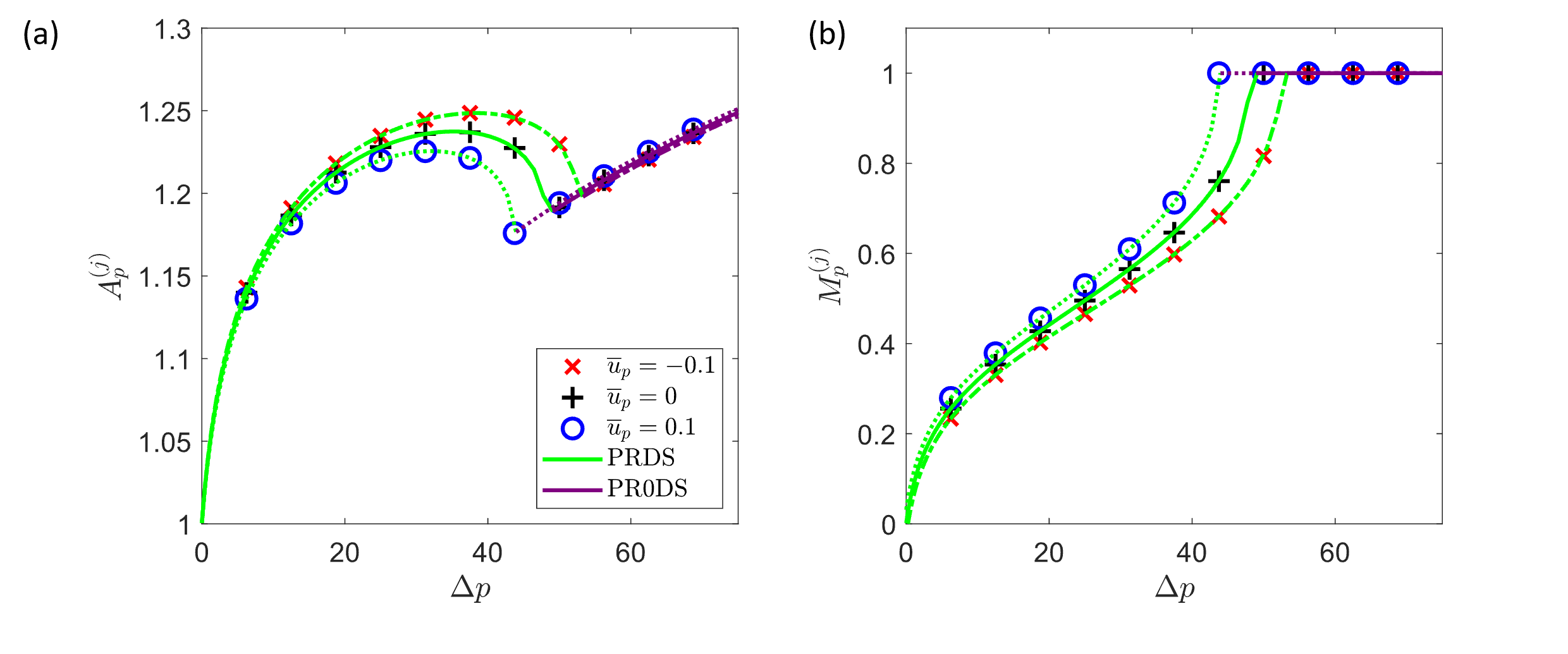}
\caption{Effect of underlying base flow through the junction . The nonlinear tube law ($m=10$, $n=3/2$) is used here.The properties of the flow on the parent side of the junction are shown as a function of the dimensionless driving pressure comparing the full numerical simulations (symbols) with solutions for the junction Riemann problem for $\overline{u}_p=0$ ($+$, solid lines), $\overline{u}_p=0.1$ ($o$, dotted lines) and $\overline{u}_p=-0.1$ ($\times$, dashed lines). (a) cross-sectional area in the parent vessel; (b) Mach number in the parent vessel}.
\label{fig:NonZeroBaseFlow}
\end{figure}

Cardiovascular networks exhibit an underlying base flow and so an arterial (venous) junction would experience net flow into the daughter vessels (parent vessel) over a cardiac cycle. In order to assess how this base flow influences the behaviour of the model with the nonlinear tube law (the linear tube law is qualitatively similar), Fig.~\ref{fig:NonZeroBaseFlow} compares the predictions of the full numerical simulations to the predictions of the Riemann problem for both positive and negative base flow, again plotting bifurcation diagrams in terms of the driving pressure amplitude showing the amplitude of the reflected wave in the parent vessel at the junction (Fig.~\ref{fig:NonZeroBaseFlow}a) and the Mach number of the flow in the parent vessel at the junction (Fig.~\ref{fig:NonZeroBaseFlow}b). The behaviour in each case is qualitatively similar to the cases reported in Fig.~\ref{fig:CompareAnalytics}. The critical driving pressure for the onset of resonance reduces as the base flow increases in speed towards the junction. Underlying flow towards the junction reduces the critical driving pressure, while underlying flow away from the junction increases the critical driving pressure. In summary, increasing the base flow through the junction promotes the onset of resonance through transition to transcritical flow.

\section{Discussion}
\label{sec:discussion}

In this study we have considered the propagation of a large-amplitude pressure wave across a symmetric blood vessel junction (from a parent vessel into two identical daughter vessels) driven by an abrupt pressure perturbation at the inlet, as a precursor to studying the propagation of large-amplitude pressure waves in cardiovascular networks. We studied this system using three complementary approaches: nonlinear simulations using a generalisation of a specifically constructed finite-volume method with explicit shock capture \cite{Brook1999}; a junction Riemann problem (\emph{e.g.} \cite{contarino2016junction}) and an analytical theory based on extending small-amplitude approaches (\emph{e.g.} \cite{Lighthill_Book,Pedleybook}) to include higher-order terms. 

We have shown that for sufficiently low driving pressures the flow across the junction is subcritical, with all three approaches predicting that a single incoming shock wave is reflected as single rarefaction wave back along the parent vessel towards the inlet, and transmitted as shock waves in each of the two daughter vessels (Figs.~\ref{fig:Fig2_LinearProfile},\ref{fig:Fig2_NonlinearProfile}); predictions of the full numerical model and the junction Riemann problem agree almost perfectly in this regime, and the analytical theories asymptote to the others in the limit of zero driving (Fig.~\ref{fig:BifurcationDiagram_NonLinear}). However, as the driving pressure increases the spatial profile of the rarefaction wave in the parent vessel expands and eventually resonates with a stationary wave at the junction. This transition occurs as the flow in the parent vessel becomes transcritical \emph{i.e.} the flow speed and the wave speed in the parent vessel become equal at the junction (Fig.~\ref{fig:BifurcationDiagram_NonLinear}c). Predictions of the critical driving pressure for the onset of resonance agree very well between the full numerical model and the junction Riemann problem, but this transition is missed entirely by the analytical theory (Fig.~\ref{fig:CompareAnalytics}). Beyond this critical point the system enters a resonant state which exhibits at least one additional shock wave in the Riemann problem; predictions of the full numerical model and the junction Riemann problem agree qualitatively in this regime, though the full numerical simulations cannot capture these additional shock waves and exhibit some numerical issues, and for even larger driving the method eventually fails to converge entirely. An obvious item of future work would be to directly include our junction Riemann problem within the fully nonlinear finite volume code, replacing the characteristic extrapolation technique which is currently used to span the junction. Such a combined approach would then be capable of examining a much more general class of wave propagation problems, and would naturally extend to a large scale network.

Furthermore, we have shown that an abrupt decrease in vessel stiffness as the parent vessel divides across the junction (as expected in most areas of the human body) promotes the onset of resonance (Fig.~\ref{fig_stiffness}). In addition, simple baseline flow driven through the junction (by a prescribed flow rate) was shown to promote (suppress) the onset of resonance when that baseline flow is directed toward (away from) the daughters as in arterial (venous) networks (Fig.~\ref{fig:NonZeroBaseFlow}).

We have shown how subcritical solutions of the junction Riemann problem can eventually become transcritical as the far-field conditions vary. However, we note that there may be other supercritical branches of solution entirely disconnected from the solutions identified here \cite{Murillo2021}, facilitating regions of parameter space where the system has multiple co-existing solutions (\emph{e.g} \cite{toro2012simplified,toro2013flow}). It should be noted that the range of far-field conditions considered here does span regimes where the initial conditions are supercritical. However, this supercriticality only plays a role in the solution for a very small wedge of the space (white region close to ${\cal A}_p=1$, ${\cal U}_p=1$ in Fig.~\ref{fig:Riemann_Profiles}c, left panel); connecting the supercritical solution in this wedge to the other solutions identified across the parameter space is an active area of future work.

The theoretical model adopted for this study is deliberately very simple, missing many of the features of human cardiovascular networks. In particular, the model employs only a very simplified description of the vessel wall mechanics, neglects vessel wall tapering and all viscous effects. The latter assumption is particularly limiting as the analysis not only misses the viscous losses associated with flow along a single vessel (which could be included using an empirical term in the governing equations \cite{Brook1999,cancelli1985separated}), it also neglects energy dissipated by the complicated three-dimensional flows generated in the neighbourhood of the junction (which can be computed \emph{e.g.} \cite{dewilde2016assessment}) and also the energy dissipated across the propagating shock fronts themselves as the vessel walls rapidly expand (see discussion in Pedley \cite{Pedleybook}). However, our model reflects the simplest version of the system which admits nonlinear shock waves, and it is an active area of future work to consider how the predictions are modified by viscous effects.

\vskip1pc

\noindent \textbf{Data Access:} This article has no experimental data. Numerical data for the figures in this paper was generated using MATLAB 2022b and can be accessed at http://dx.doi.org/10.5525/gla.researchdata.1560. \\

\noindent \textbf{Author Contributions:} TAS carried out the full numerical simulations and the modified linear theory and drafted the manuscript; ISO carried out the analysis of the junction Riemann problem and drafted the manuscript; PSS and DM conceived and coordinated the study, contributed to developing the numerical codes and edited the manuscript. All authors gave final approval for publication. \\

\noindent \textbf{Competing Interests:} We declare we have no competing interests. \\

\noindent \textbf{Funding:} TAS and PSS acknowledge funding from EPSRC grant no.~EP/P024270/1. ISO acknowledge funding from the Petroleum Technology Development Fund (PTDF). PSS acknowledges funding from EPSRC grant nos.~EP/N014642/1, EP/S020950/1 and EP/S030875/1. \\

\noindent \textbf{Acknowledgements:} Helpful discussions with B.S.~Brook (University of Nottingham, UK), N.A.~Hill (University of Glasgow, UK), E.F.~Toro, L.~Muller and A.~Siviglia (all University of Trento, It) are very gratefully acknowledged. \\

\appendix

\section{Small amplitude waves}
\label{App_PertTheory}

In this appendix we extend the classical small-amplitude analysis for flow in elastic-walled blood vessels \cite{Lighthill_Book,Pedleybook} to derive approximations to the amplitudes of the reflected and transmitted waves at the junction. The derivation assumes that the system is perturbed by a small amplitude increase in pressure at the upstream end of the parent vessel, denoted $\Delta p\ll 1$, which propagates uniformly toward the junction where it is partially reflected back into the parent vessel, and partly transmitted into the daughter vessels. Note that (as in the main text) we assume all vessels have the same baseline cross-sectional area, but the parent and daughter vessels can exhibit a difference in their elastic stiffness.

For the linear tube law (Eqs.~(\ref{eq:GovEqu_Pressure}),(\ref{eq:tubelaw}) with $m=1$ and $n=0$), we calculate the corresponding perturbation in cross-sectional area at the upstream end of the parent vessel as $\Delta A=\Delta p/k_p=\Delta p$.

We denote the corresponding profiles of the cross-sectional area and flow rate in each vessel as ($i=p,d$)
\begin{subequations}
\label{Eq_Perturbation}
\begin{align} 
A_i=1 + \hat{A}_i^{(1)}\Delta A + \hat{A}_i^{(2)}\Delta A^2 + O(\Delta A^3), \\
q_i=0 + \hat{q}_i^{(1)}\Delta A + \hat{q}_i^{(2)}\Delta A^2 + O(\Delta A^3), 
\end{align}
\end{subequations}
where hatted quantities are to be determined. The corresponding pressure in each vessel can be approximated using the linear tube law (Eqs.~(\ref{eq:GovEqu_Pressure}),(\ref{eq:tubelaw}) with $m=1$ and $n=0$), so that
\begin{align}
p_i = k_i\hat{A}_i^{(1)}\Delta A + k_i\hat{A}_i^{(2)}\Delta A^2 + O(\Delta A^3).
\end{align}

The governing equations of mass and momentum (Eqs.~\ref{eq:GovEqu_Mass},\ref{eq:GovEqu_Momentum}) can be solved to first and second-order in perturbation amplitude. 

The solution at first-order is a classical derivation that will be outlined here briefly \cite{Pedleybook}. Firstly, we substitute the ansatz in Eq.~(\ref{Eq_Perturbation}) into the mass and momentum governing equations (Eqs.~(\ref{eq:GovEqu_Mass}), (\ref{eq:GovEqu_Momentum})) retaining terms at $O(\Delta A)$. By cross-differentiating we reduce to the classical wave equation in each vessel expressed in terms of the vessel cross-sectional area with corresponding wavespeed $c_i = \sqrt{k_i}$ ($i=p,d$). Hence, the resulting wave profiles can be written as ($i=p,d$)
\begin{align}
\hat{A}^{(1)}_i & =\frac{1}{\sqrt{k_i}}\left( f^{(1)}_{i}(x-c_it)+g^{(1)}_{i}(x+c_it) \right)  \label{WaveSol_2} \\
\hat{q}^{(1)}_i & =f^{(1)}_{i}(x - c_it)-g^{(1)}_{i}(x+c_it), \label{WaveSol_1}
\end{align}
in terms of wave profiles $f^{(1)}_i(z)$ and $g^{(1)}_i(z)$ in each vessel to be determined. In what follows we approximate all the propagating waves as square waves with spatially uniform profiles on either side of the propagating wavefront, which can be expressed mathematically in terms of a Heaviside function defined as
\begin{equation}
H(z) =1, \qquad (z<0), \qquad H(z) = 0, \qquad (z>0).
\end{equation}
Flow in the parent vessel is driven by the inlet condition, which then sets the wave profile propagating toward the junction in the form $f^{(1)}_p(z) = H(z)$. Furthermore, all information in the daughter vessels is propagating away from the junction so that $g^{(1)}_d(z)=0$. After the incoming wave has encountered the junction the remaining profiles can be computed by applying the  junction boundary conditions (Eqs.~(\ref{Eq_Junc_Mass}), (\ref{Eq_Junc_TotPres})) to obtain
\begin{align}
g^{(1)}_p(z) &\equiv \overline{g}^{(1)}_p H(z), & \overline{g}^{(1)}_p&=\frac{\sqrt{k_d}-2\sqrt{k_p}}{2+\sqrt{k_d}},  \\
f^{(1)}_d(z) &\equiv \overline{f}^{(1)}_d H(z), & \overline{f}^{(1)}_d&=\frac{1+\sqrt{k_p}}{2+\sqrt{k_d}}.
\end{align}

To improve this approximation we continue to the following order in perturbation amplitude. Following the same approach and now retaining terms of order $\Delta A^2$ gives in each vessel ($i=p,d$)
\begin{align}
\frac{\partial^{2}\hat{A}_i^{(2)}}{\partial t^{2}}-k_i\frac{\partial^{2}\hat{A}_i^{(2)}}{\partial x^{2}} & =\frac{\partial^{2}\left[\hat{q}_i^{(1)}\right]^{2}}{\partial x^{2}}+\frac{k_i}{2}\frac{\partial^{2}\left[\hat{A}_i^{(1)}\right]^{2}}{\partial x^{2}},\label{eq:Perfect_GE}
\end{align}
To the best of our knowledge this system cannot be solved analytically without further assumptions. To make analytical progress we assume that the local spatial derivatives are zero (which will be true away from the propagating wavefronts). In this case all forcing terms are zero and the second-order solution also follows a linear wave equation with wavespeed $c_i$ and so solutions can be written as ($i=p,d$)
\begin{align}
\hat{A}^{(2)}_i & =\frac{1}{\sqrt{k_i}}\left(f^{(2)}_{i}(x-c_it )+g^{(2)}_{i}(x+c_it) \right),  \label{WaveSol_2_2} \\
\hat{q}^{(2)}_i & =f^{(2)}_{i}(x-c_it)-g^{(2)}_{i}(x+c_it), \label{WaveSol_1_2}
\end{align}
in terms of wave profiles $f^{(2)}_i(z)$ and $g^{(2)}_i(z)$ in each vessel to be determined. Since there is no incoming perturbation at this order, all contributions are generated through interaction with the first-order flow at the junction and so $f^{(2)}_p=0$ and $g^{(2)}_d=0$. Assuming the constituent waves are again square waves, applying the junction boundary conditions (Eqs.~\ref{Eq_Junc_Mass}, \ref{Eq_Junc_TotPres}) the solution of the system takes the form
\begin{align}
{g}_p^{(2)}(z)&=\overline{g}_p^{(2)}H(z), & \overline{g}_p^{(2)}&=\frac{-3}{(2 \sqrt{k_p}+\sqrt{k_d})} \left(\frac{1+\sqrt{k_p}}{2+\sqrt{k_d}}\right)^2,\\
{f}_d^{(2)}(z)&=\overline{f}_d^{(2)}H(z), & \overline{f}_d^{(2)}&=-\frac{1}{2} \overline{g}_p^{(2)}.
\end{align}
These amplitudes are substituted back into the ansatz for the vessel cross-sectional area and flux (Eq.~\ref{Eq_Perturbation}) to obtain first- and second-order predictions for the amplitudes of the transmitted and reflected waves, as listed in Eq.~(\ref{MLP}), where $k_p$ has been set to 1 in accordance with our non-dimensionalisation.

\bibliographystyle{unsrt} 
\bibliography{References} 

\end{document}